\documentclass[sigconf]{acmart}

\usepackage{listings}
\usepackage{xcolor}
\usepackage{multirow}
\usepackage{url}
\usepackage{threeparttable}
\usepackage{balance}

\usepackage{pifont}
\newcommand{\xmark}{\ding{55}}%

\usepackage{listings}
\usepackage{xcolor} 
\usepackage{multirow}
\usepackage{booktabs} 
\usepackage{amsmath}

\usepackage{graphicx}
\usepackage[belowskip=-12pt]{caption}
\usepackage{subcaption}
\usepackage{rotating}
\usepackage{tabularx}
\usepackage[noend,linesnumbered, ruled]{algorithm2e}
\usepackage{wrapfig}
\usepackage{listings}
\usepackage{color, colortbl}
\usepackage{balance} 
\usepackage{lscape}
\SetKw{KwStep}{Step}
\usepackage{hyperref}
\usepackage{xspace}
\usepackage{framed}
\usepackage{syntax}
\usepackage{comment}

\usepackage{caption} 
\usepackage{caption} 
\usepackage[framemethod=TikZ]{mdframed}
\usepackage{soul}
\usepackage{flushend}
\usepackage{enumitem}

\usepackage{tikz}

\definecolor{codegreen}{rgb}{0,0.6,0}
\definecolor{codegray}{rgb}{0.5,0.5,0.5}
\definecolor{codepurple}{rgb}{0.58,0,0.82}
\definecolor{backcolour}{rgb}{0.95,0.95,0.92}
\definecolor{Gray}{gray}{0.1}
\lstdefinestyle{mystyle}{
	backgroundcolor=\color{backcolour},   
	commentstyle=\color{codegreen},
	keywordstyle=\color{magenta},
	numberstyle=\tiny\color{codegray},
	stringstyle=\color{codepurple},
	basicstyle=\scriptsize,
	breakatwhitespace=false,         
	breaklines=true,                 
	captionpos=b,                    
	keepspaces=true,                 
	numbers=left,                    
	numbersep=5pt,                  
	showspaces=false,                
	showstringspaces=false,
	showtabs=false,                  
	tabsize=2
}

\lstdefinelanguage{Pythonna}{%
	language     = python,
	morekeywords = {to_categorical, flow_from_directory, pad_sequences, load_image}
}

\lstdefinestyle{customc}{
	belowcaptionskip=1\baselineskip,
	breaklines=false,
	frame= single,
	breaklines = true,
	xleftmargin=\parindent,
	language= Pythonna,
	showstringspaces=false,
	basicstyle=\footnotesize\ttfamily,
	keywordstyle=\bfseries\color{green!40!black},
	commentstyle=\itshape\color{purple!40!black},
	identifierstyle=\color{blue},
	stringstyle=\color{codegreen},
	backgroundcolor=\color{gray!4}
}

\graphicspath{{./figures/}}

\newcommand{\tabref}[1]{Table~\ref{#1}}

\newcommand{\lstref}[1]{Listing~\ref{#1}}

\newcommand{\secref}[1]{\S\ref{#1}}

\newcommand{\etal}{{\em et al.}\xspace}

\newcommand{\sof}{\textit{Stack Overflow}\xspace}
\newcommand{\keras}{\textit{Keras}\xspace}

\newcommand{\gh}{GitHub\xspace}

\newcounter{rqs}
\stepcounter{rqs}

\newcounter{NumObservations}
\stepcounter{NumObservations}
\definecolor{shadecolor}{rgb}{.9,.9,.9}

\def\BibTeX{{\rm B\kern-.05em{\sc i\kern-.025em b}\kern-.08em
    T\kern-.1667em\lower.7ex\hbox{E}\kern-.125emX}}

\AtBeginDocument{%
  \providecommand\BibTeX{{%
    \normalfont B\kern-0.5em{\scshape i\kern-0.25em b}\kern-0.8em\TeX}}}

\begin{document}

\title {
An Effective Data-Driven Approach for Localizing Deep Learning Faults
}





\author{Mohammad Wardat}
\email{wardat@iastate.edu}
\affiliation{%
	\institution{Dept. of Computer Science, Iowa State University}
	\streetaddress{226 Atanasoff Hall}
	\city{226 Atanasoff Hall, Ames}
	\state{IA}
	\postcode{50010}
	\country{USA}
}

\author{Breno Dantas Cruz}
\email{bdantasc@iastate.edu}
\affiliation{%
	\institution{Dept. of Computer Science, Iowa State University}
	\streetaddress{226 Atanasoff Hall}
	\city{226 Atanasoff Hall, Ames}
	\state{IA}
	\postcode{50010}
	\country{USA}
}

\author{Wei Le}
\email{weile@iastate.edu}
\affiliation{%
	\institution{Dept. of Computer Science, Iowa State University}
	\streetaddress{226 Atanasoff Hall}
	\city{226 Atanasoff Hall, Ames}
	\state{IA}
	\postcode{50010}
	\country{USA}
}

\author{Hridesh Rajan}
\email{hridesh@iastate.edu}
\affiliation{%
	\institution{Dept. of Computer Science, Iowa State University}
	\streetaddress{226 Atanasoff Hall}
	\city{226 Atanasoff Hall, Ames}
	\state{IA}
	\postcode{50010}
	\country{USA}
}




\begin{abstract}
Deep Learning (DL) applications are being used to solve problems in critical domains (e.g., autonomous driving or medical diagnosis systems). Thus, developers need to debug their systems to ensure that the expected behavior is delivered. However, it is hard and expensive to debug DNNs. When the failure symptoms or unsatisfied accuracies are reported after training, we lose the traceability as to which part of the DNN program is responsible for the failure. Even worse, sometimes, a deep learning program has different types of bugs. To address the challenges of debugging DNN models, we propose a novel data-driven approach that leverages model features to learn problem patterns. Our approach extracts these features, which represent semantic information of faults during DNN training. Our technique uses these features as a training dataset to learn and infer DNN fault patterns. Also, our methodology automatically links bug symptoms to their root causes, without the need for manually crafted mappings, so that developers can take the necessary steps to fix faults. We evaluate our approach using real-world and mutated models. Our results demonstrate that our technique can effectively detect and diagnose different bug types. Finally, our technique achieved better accuracy, precision, and recall than prior work for mutated models. Also, our approach achieved comparable results for real-world models in terms of accuracy and performance to the state-of-the-art. 
\end{abstract}


\maketitle
\section{Introduction}
\label{sec:intro}
Nowadays, Deep Learning (DL) is becoming pervasive in our daily lives. Developers are leveraging DL-based applications to solve critical tasks, such as fraud detection~\cite{yaram2016machine}, medical diagnosis~\cite{janowczyk2016deep, miotto2018deep}, facial recognition~\cite{yang2021larnet}, and autonomous driving~\cite{tian2018deeptest}. However, just like traditional software, DL-based applications are also susceptible to bugs. However, the process of debugging such applications remains hard and expensive. Requiring developers to be experts on DNN bug behavior when using DL-based frameworks (e.g., \textit{TensorFlow}~\cite{TensorFlow} and \keras~\cite{Keras}). For example, TensorBoard~\cite{mane2015tensorboard} is a tool that enables developers to visualize the model behavior during training. However, TensorBoard does not state which hyperparameters are causing the fault, thus limiting its usefulness for non-experts.

To address this problem, recent software engineering research efforts have focused on two directions (i.e., DNN fault localization~\cite{zhang2020detecting, wardat21DeepLocalize} and repair~\cite{wardat2021deepdiagnosis, ZhangAUTOTRAINER}). Fault localization focuses on identifying and localizing bugs based on sets of rules (e.g., loss not decreasing). DNN repair approaches leverage bug symptoms to diagnose their root causes and provide corresponding actionable fixes. Nevertheless, both directions rely on scenarios with one bug type. That is, a root cause with corresponding symptoms manifesting at training time. However, real-world models may have several bug types, each having different symptoms and root causes~\cite{islam19, ZhangAUTOTRAINER, wardat2021deepdiagnosis}. 
Prior works have supported a few types of bugs at once~\cite{schoop2020scram, zhang2020detecting, wardat2021deepdiagnosis, wardat21DeepLocalize}. For example, AUTOTRAINER cannot simultaneously handle the loss function and dropout rate due to the complexity of solving multiple root causes~\cite{ZhangAUTOTRAINER}. 
Furthermore, it is difficult to accurately localize multiple root causes only by observing their symptoms~\cite{islam19, zhang2018empirical}. Prior works have focused on detecting the first symptom appearing during training, (e.g., unchanged weight), thus restricting the supported fault types~\cite{schoop2020scram, zhang2020detecting, wardat2021deepdiagnosis, wardat21DeepLocalize}. 

To fill this gap, we introduce Deep4Deep, a learning-driven approach that builds over prior research on program analysis for fault localization ~\cite{wardat2021deepdiagnosis, ZhangAUTOTRAINER}. Also, Deep4Deep is inspired by data-driven approaches to solving complex problems (e.g., image classification~\cite{ciregan2012multi},
natural language pricessing~\cite{sutskever2014sequence}, speech recognition~\cite{hinton2012deep}, and intrusion detection~\cite{du2017deeplog}). SE research is currently investigating how to leverage such approaches to issues related to code intelligence tasks~\cite{buch2019learning, li2019improving, bielik2016phog, gupta2017deepfix, chen2018tree, wei2019code, allamanis2016convolutional}. These issues require extracting large training datasets from a program. Lastly, data-driven approaches can be extended to support bug detection problems in DL software while reducing the cost of debugging DL software~\cite{balog2017deepcoder}. Our prototype allows developers to detect new bug types, which may emerge over time, by finetuning with a dataset containing the new bug types.

For the design of our approach, we followed the intuition that a Long Short-Term Memory (LSTM) model~\cite{hochreiter1997long} will facilitate the process of mapping symptoms to their root causes. Such model architecture can learn the complex relationship between symptoms to their root causes. Furthermore, this architecture would remain consistent with the addition of emerging new faults in the real world. To do so, our proposed technique automatically learns the relation between root causes and their symptoms. Therefore, Deep4Deep reduces the effort while improving the scalability of debugging DNNs. We propose a powerful representation of the DL model to enable the automatically learn of semantic features from dynamic and static sources~\cite{zhang2019novel,abad2019supporting, Jialun2022DeepFD, rahman2013and}. In particular, for dynamic analysis
Deep4Deep analyzes the dynamic values of each parameter (i.e., weights and metrics) at a timestamp, which enables a holistic view of model behavior during training. This holistic view allows for the simultaneous observation of multiple symptoms. For static features, Deep4Deep extracts static information from the model source code. In particular, it extracts token vectors from the model in the ONNX format. By integrating these dynamic and static features, our design strategy can capture semantic distinctions between programs while building highly accurate prediction models~\cite{rahman2013and}.

To summarize, this paper makes the following contributions:
\begin{itemize}
    \item We formulate the fault localization of a deep learning program as a machine learning problem. We identify a set of challenges for encoding DNN fault localization as a sequence-to-sequence supervised learning problem.
    \item We present a data-driven approach, Deep4Deep, which learns from the semantic features (i.e., dynamic and static) that can represent a DNN model behavior at training time. 
    \item We develop an approach for systematically generating training datasets for supervised learning. The training datasets consist of models with multiple bugs. We hope that our dataset can help other researchers validate their debugging and repair tools for multi-bug detection.  
    \item We extensively evaluate Deep4Deep in both the real world (i.e., from StackOverflow and GitHub) and mutated benchmarks. Our results show that Deep4Deep can simultaneously handle 8 types of bugs while correctly diagnosing their root causes. Deep4Deep offers higher precision, recall, and accuracy than existing program analysis techniques.
\end{itemize}

The rest of the paper is organized as follows: 
In \secref{sec:background} we motivate our work by providing an example. 
In \secref{sec:Approach} we describe Deep4Deep, our DL-based approach for DNN bug localization.
In \secref{sec:EVALUATION} we provide the evaluation of our approach, and \secref{sec:THREATSTOVALIDITY} we discuss the threats to validity.
In \secref{sec:relatedwork} we discuss the related works. Lastly, in \secref{sec:conclusion} we provide the concluding remarks and discuss future work directions.

\begin{figure*}[ht!]
	\centering
	\includegraphics[width=\linewidth,trim={0cm 0cm 0cm 0cm}]{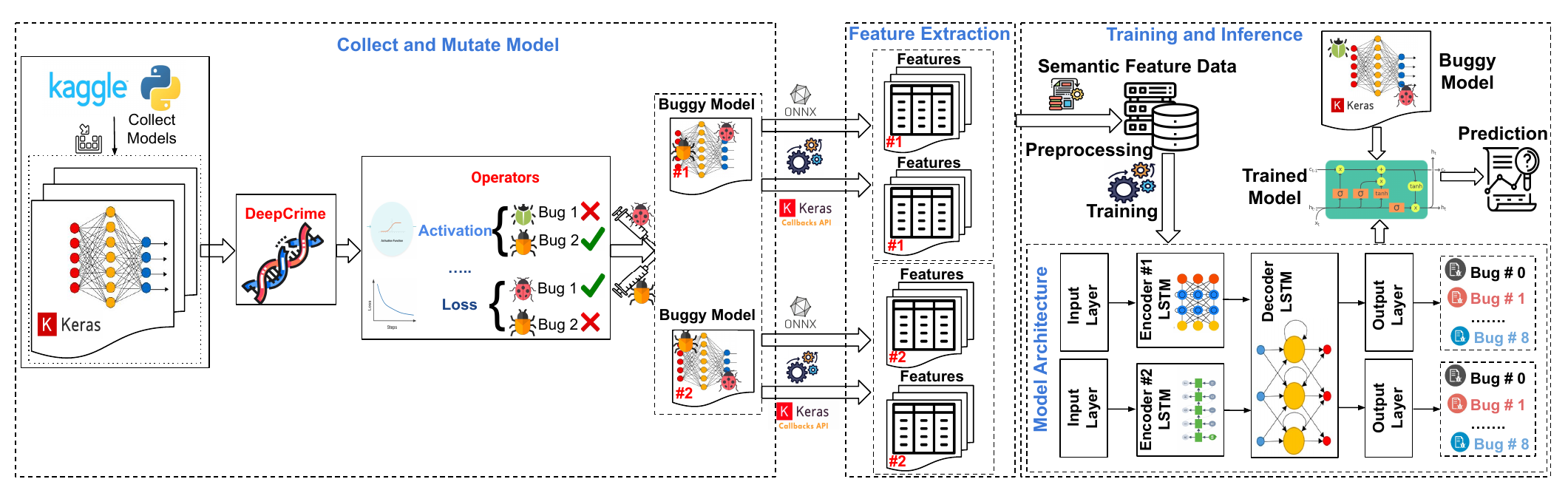}
    \caption{Overview of Deep4Deep.}
	\label{Process}
\end{figure*}

\section{A Motivating Example}
\label{sec:background}
In this section, we motivate our work by providing an example that illustrates the challenge of bug localization in DL-based programs via machine learning.



\begin{table}[!htb]
\caption{Comparison between different approaches}
\label{tab:results-operators}
\scalebox{0.625}{
\begin{tabular}{|l|c|c|c|c|c|c|}
\hline
 \cellcolor[HTML]{EFEFEF} \textbf{Hyperparameter}                & \cellcolor[HTML]{EFEFEF} \textbf{UMLAUT} & \cellcolor[HTML]{EFEFEF} \textbf{Auto} & \cellcolor[HTML]{EFEFEF} \textbf{DeepLoc} & \cellcolor[HTML]{EFEFEF} \textbf{DeepDiag} &\cellcolor[HTML]{EFEFEF}  \textbf{DeepFD} &\cellcolor[HTML]{EFEFEF}  \textbf{D4D}\\ \hline
\textbf{Loss\_Function}          & \xmark              & \xmark            & \checkmark              & \checkmark               & \checkmark             & \checkmark\\ \hline
\textbf{Batch\_Size}    \cellcolor[HTML]{EFEFEF}          & \cellcolor[HTML]{EFEFEF} \xmark              & \cellcolor[HTML]{EFEFEF} \checkmark           & \cellcolor[HTML]{EFEFEF} \xmark               & \cellcolor[HTML]{EFEFEF} \xmark                & \cellcolor[HTML]{EFEFEF} \xmark            & \checkmark \cellcolor[HTML]{EFEFEF} \\ \hline
\textbf{Learning\_Rate}          & \checkmark             & \checkmark           & \checkmark              & \checkmark               & \checkmark           & \checkmark  \\ \hline
\textbf{Activation\_Function} \cellcolor[HTML]{EFEFEF}    & \cellcolor[HTML]{EFEFEF} \checkmark             & \cellcolor[HTML]{EFEFEF} \checkmark           & \cellcolor[HTML]{EFEFEF} \checkmark              &\cellcolor[HTML]{EFEFEF}  \checkmark               &  \cellcolor[HTML]{EFEFEF} \checkmark           & \cellcolor[HTML]{EFEFEF} \checkmark  \\ \hline
\textbf{Optimisation\_Function}  & \xmark              & \checkmark           & \xmark               & \checkmark               & \checkmark            & \checkmark \\ \hline
\textbf{Gradient\_Clip}      \cellcolor[HTML]{EFEFEF}     & \cellcolor[HTML]{EFEFEF} \xmark              &  \cellcolor[HTML]{EFEFEF} \checkmark           & \cellcolor[HTML]{EFEFEF} \xmark               & \cellcolor[HTML]{EFEFEF} \xmark                &\cellcolor[HTML]{EFEFEF}  \xmark             & \cellcolor[HTML]{EFEFEF} \checkmark \\ \hline
\textbf{Weights\_Initialisation} & \xmark              & \checkmark           & \xmark               & \checkmark               & \xmark           & \checkmark   \\ \hline
\textbf{Dropout\_Rate}    \cellcolor[HTML]{EFEFEF}        & \cellcolor[HTML]{EFEFEF} \checkmark             & \cellcolor[HTML]{EFEFEF} \xmark            & \cellcolor[HTML]{EFEFEF} \xmark               & \cellcolor[HTML]{EFEFEF} \xmark                & \cellcolor[HTML]{EFEFEF} \xmark              & \cellcolor[HTML]{EFEFEF}\checkmark\\ \hline
\end{tabular}
}

\end{table}


\begin{table}[!htb]
    \centering
    \caption{Result from Motivating Example}
    \label{tab:Result-motivate}
\scalebox{0.88}{
\begin{tabular}{|c|l|}
\hline
\rowcolor[HTML]{EFEFEF} 
\textbf{Approach}               & \multicolumn{1}{c|}{\cellcolor[HTML]{EFEFEF}\textbf{Output}} \\ \hline
\textbf{UMLAUT}        & No Output                                                    \\ \hline
\rowcolor[HTML]{EFEFEF} 
\textbf{DeepLocalize}  & Layer 7: Error in forward                                    \\ \hline
\textbf{AUTOTRAINER}   & Training problems: {[}'not\_converge'{]}                     \\ \hline
\rowcolor[HTML]{EFEFEF} 
\textbf{DeepDiagnosis} & Layer 7: Error in forward, Change loss function              \\ \hline
\textbf{Deep4Deep}     & class 1 - loss function, class 7 - optimizer                 \\ \hline
\end{tabular}
}
\end{table}


Consider the code snippet in Listing~\ref{lst:motivation}, a simplified version of the code from \keras~\cite{simplemnistconv}, which contains two root causes in the 
MNIST model, leading to low performance. This program defines a model to solve a classification problem using the MNIST dataset. In this program, Line 9 has two faults. First, the loss function used \texttt{mean\_square\_error}, instead of \texttt{categorical\_crossentropy}. Second,  the optimizer \texttt{sgd} is used instead of \texttt{Adam}. In a classification problem~\cite{tuske2015integrating}, a DNN program typically is composed of a combination of layers, such as {\it convolutional}, {\it pooling}, {\it fully connected}, and {\it softmax}. The cross-entropy loss is commonly used after a softmax layer to normalize the network's output. Several adaptive optimizers were developed to automatically update the weights by adding the error derivatives to the previous weights. Adaptive optimizers, such as Adam~\cite{kingma2014adam}, simplify learning rate settings and lead to faster convergence when compared to other optimizers~\cite{defossez2020convergence}.
\begin{lstlisting}[language = python,label={lst:motivation}, caption={ Simple Keras MNIST Model~\cite{simplemnistconv}}]
model = Sequential([Input(shape=input_shape),
        Conv2D(32, kernel_size=(3, 3), activation="relu"),
        MaxPooling2D(pool_size=(2, 2)),
        Conv2D(64, kernel_size=(3, 3), activation="relu"),
        layers.MaxPooling2D(pool_size=(2, 2)),
        Flatten(),
        Dropout(0.5),
        Dense(num_classes, activation="softmax")])
model.compile(loss="MSE", optimizer="sgd", metrics=["acc"])
model.fit(x_train, y_train, batch_size=128, epochs=15)
\end{lstlisting}~\label{listing}


Table~\ref{tab:results-operators} shows a comparison of Deep4Deep with the most relevant SoTA techniques capable of detecting bugs in DNN programs. In the first column, we list a set of hyperparameters that may lead to mistakes. In the table, a ``\checkmark'' represents the supported bug diagnosis while a ``\xmark'' indicates the unsupported one. Specifically, DeepDiagnosis does not support bugs related to dropout rate~\cite{wardat2021deepdiagnosis}, AUTOTRAINER~\cite{ZhangAUTOTRAINER} does not handle faults related to the loss function. Also, DeepLocalize~\cite{wardat21DeepLocalize} only supports bugs related to numerical errors (i.e., INF, NaN, etc.). UMLAUT~\cite{schoop2021umlaut} exclusively supports classification problems. 

Table~\ref{tab:Result-motivate} shows the results reported by the four SoTA tools for Listing~\ref{lst:motivation}. For each tool, we follow the instructions available in their reproducibility packages (i.e., DeepLocalize~\cite{myRepo}, DeepDiagnosis~\cite{DDRepo}, AUTOTRAINER~\cite{AUTORepo} and UMLAUT~\cite{UMLAUTRepo}) and carefully made semantic changes to their respective code to enable the analysis of the code in~\lstref{lst:motivation}. We validated all changes by comparing their analysis results against the ones of their respective benchmarks. The results show that the SoTA tools do not support all bug types in the target DNN program. Also, a few of them did not identify any bugs. In particular, UMLAUT finished the training after approximately 700 seconds without indicating any bugs. DeepLocalize indicated a problem in the last layer after approximately 20 seconds. However, it did not provide any information on how to fix it. DeepDiagnosis returned the following message after approximately 38 seconds: ``Batch 4 Layer 7: Error in forwarding terminating training.'' This message indicates a batch error in the forwarding propagation stage during training. DeepDiagnosis provided a message with the correct fix suggestion for the loss function, but it missed the second bug. AUTOTRAINER was able to fix the bug in the optimizer by changing it from \texttt{sdg} to \texttt{Adam}, but missed the incorrect usage of the loss function. After 1140 seconds, AUTOTRAINER did not completely solve the problem or provided diagnostic information for all bugs at once.

Our approach Deep4Deep correctly reported faults in the program after approximately 2936 seconds. Also, Deep4Deep returned the bug classes, (i.e., \textbf{class 1 - loss function} and \textbf{class 7 - optimizer}), which can guide developers to perform accurate fixes to their program (more details in~\secref{sec:Approach}). Even though Deep4Deep takes longer, it can provide fault localization for all bug types listed in Table~\ref{tab:results-operators}. Since SoTA approaches use hard-coded rules for bug detection, it is hard to ensure their consistency when adding support to new bug types. 
Additional rules have to be verified to avoid breaking prior ones. Our data-driven approach learn bug patterns from data. These patterns can help detect new bug types, 
as we discuss in Section~\ref{sec:DFDVSD4D}.

\section{Approach}
\label{sec:Approach}

\begin{table*}[!htb]
     \setlength{\parskip}{.01cm}
    \setlength{\belowcaptionskip}{.05cm}

  \caption{Description of the 15 DNN dynamic features used by Deep4Deep for Fault Localization}~\label{feature}
\scalebox{0.825}{
\begin{tabular}{|l|l|l|p{41.em}|}
\hline
\textbf{No} & \textbf{Feature   Names}            & \textbf{Symbols} & \textbf{Description}                                                                                                                                                                    \\ \hline

\rowcolor[HTML]{EFEFEF} 1           & Loss $\mid$ Accuracy & LS $\mid$ AC          & Stores the loss and accuracy in each iteration                                                                                                                                           \\ \hline
2           & Data Range                          & DR               & Computes the maximum and minimum value of the training dataset.     \\\hline

\rowcolor[HTML]{EFEFEF} 3           & Mean \{Weights$\mid$ Gradients$\mid$ Biases\} & M\_\{W$\mid$ G$\mid$ B\}   & Computes the Input’s mean value.                                                                                                                                                        \\ \hline
4           & Min \{Weights$\mid$ Gradients$\mid$ Biases\} & Mi\_\{W$\mid$ G$\mid$ B\}   & Computes the Input’s minimum value.                                                                                                                                                        \\ \hline
\rowcolor[HTML]{EFEFEF} 5           & Max \{Weights$\mid$ Gradients$\mid$ Biases\} & Ma\_\{W$\mid$ G$\mid$ B\}   & Computes the Input’s maximum value.                                                                                                                                                        \\ \hline
6           & Median \{Weights$\mid$ Gradients$\mid$ Biases\} & Me\_\{W$\mid$ G$\mid$ B\}   & Computes the Input’s median value.                                                                                                                                                        \\ \hline
\rowcolor[HTML]{EFEFEF} 7           & VAR \{Weights$\mid$ Gradients$\mid$ Biases\}  & V\_\{W$\mid$ G$\mid$ B\}   & Computes the Input’s variance value.                                                                                                                                                    \\ \hline
8           & STD \{Weights$\mid$ Gradients$\mid$ Biases\}  & S\_\{W$\mid$ G$\mid$ B\}   & Computes the Input’s standard deviation value.                                                                                                                                          \\ \hline
\rowcolor[HTML]{EFEFEF} 9           & Sem \{Weights$\mid$ Gradients$\mid$ Biases\}  & Se\_\{W$\mid$ G$\mid$ B\}   & Computes the Input’s standard error of the mean value.                                                                                                                                          \\ \hline
10           & Skew \{Weights$\mid$ Gradients$\mid$ Biases\}  & Sk\_\{W$\mid$ G$\mid$ B\}   & Computes a distortion that deviates from the symmetrical bell curve, or normal distribution, in a set of data.                                                                                                                                          \\ \hline
\rowcolor[HTML]{EFEFEF} 11           & Norm \{Weights$\mid$ Gradients\}         & N\_\{W$\mid$ G\}      & Computes the norm of the gradients $\mid$ weights for each   layer.                                                                                                                          \\ \hline
                                                                                                                    \hline
12           & Vanishing Gradient                  & VG               & Computes the mean of the gradients’ absolute values.                                                                                                                                    \\ \hline
\rowcolor[HTML]{EFEFEF} 13           & Dead Node                           & DN               & Computes the percentage of inactive nodes by counting how many   inactive nodes dropped below the Threshold.                                                                            \\ \hline
14           & Saturated Activation                & SA               & Computes the percentage of total active nodes by counting the activity of ones with activation function either greater than a Max\_Threshold or less than a Min\_Threshold. \\ \hline
\rowcolor[HTML]{EFEFEF} 15          & Tune Learning                       & TL               & For each layer, computes the ratio between the norm for its corresponding gradient and weights.                                                                                                 \\ \hline
\end{tabular}
}
\label{feaures}
\begin{tablenotes}
    \small
    \item This table shows the parameters and descriptive statistics definitions. We borrow these methods for feature extraction from prior works~\cite{SageMaker, LearningRate, wardat2021deepdiagnosis, ZhangAUTOTRAINER, wardat21DeepLocalize, Jialun2022DeepFD} \\[1ex]
    
    \end{tablenotes}

\end{table*}

In this section, we present the details of our approach, Deep4Deep, for detecting bugs in DNN programs. Specifically, we detail how we automatically extract features during model training. Also, we describe the design of a Sequence to Sequence (Seq2Seq) model for fault localization. 

\subsection{Overview}
Figure~\ref{Process} shows the overview of Deep4Deep. In the first step, we prepare the training dataset by collecting the real-world models from {\it Kaggle}~\cite{Kaggle} and systematically mutating each one using DeepCrime~\cite{humbatova2021deepcrime}. Section~\ref{TrainingDataset} provides the details of the dataset creation. For feature extraction, we use both static and dynamic analysis techniques to extract those features. In particular, we use program analysis techniques~\cite{ZhangAUTOTRAINER,wardat2021deepdiagnosis} and callbacks~\cite{wardat2021deepdiagnosis, wardat21DeepLocalize} to monitor model training and collect features that likely identify the symptoms of buggy models (details in~\ref{FeatureExtraction}). Also, we use the model's static information 
after converting it to ONNX format.  Inspired by prior work~\cite{du2017deeplog, hochreiter1997long}, we developed an LSTM model that takes in sequential data as features to identify correlations between specific features and the bug root causes (additional details in~\ref{NetworkBuilding}). The output of Deep4Deep is a report that includes the probability of each root cause that can result in low performance. 
We encode the list of potential root causes as a set of labels in DNN programs (see Table~\ref{tab:operators}).

\subsection{Feature Extraction}
\label{FeatureExtraction}


Analyzing traditional programs, researchers have used Abstract Syntax Trees (ASTs)~\cite{tu2014localness} to automatically extract syntax information from source code and capture programming patterns~\cite{nguyen2015graph, nguyen2009graph}. 
Prior works have demonstrated the effectiveness of this technique in representing hidden defects~\cite{zhang2019novel, wang2016automatically}. Also, dynamic analysis can help extract such semantic features~\cite{abad2019supporting, rahman2013and}, which can distinguish between faults~\cite{wang2016automatically}. Here, we apply static and dynamic analysis to extract semantic features from DL models to detect faults. 

\textbf{Static feature extraction:}
We leverage the Open Neural Network Exchange (ONNX)~\cite{ONNX} model data to extract the model's  features. ONNX is an open-source platform that supports deep learning, and traditional machine learning models, and enables framework interoperability. ONNX defines the computation graph and operators for such models. We utilize the \textit{JSON.parse()} to extract syntactic information from JSON files. Specifically, for each model, we obtain a vector of tokens (i.e., input, name, opType, and output) of the four categories. We convert this vector to a numerical format, as the LSTM does not support text input. We encode the token sequences as a set of integers via the 'fit\_on\_texts' method. We use these sequences to build a mapping between integers and tokens. Each token, which includes ``input,'' ``name,'' ``opType,'' and ``output,'' is assigned a unique integer identifier. Because the token sequences can have different lengths, the resulting sequences may also differ in size. To ensure vector length consistency, we append zeros to the shorter ones by using the 'pad\_sequences' method. This procedure does not affect the meaning of the sequence or the LSTM layer output~\cite{wang2016automatically}, however, it is required to ensure that they can be processed by the model.

\textbf{Dynamic feature extraction:} we follow a similar procedure from prior works~\cite{wardat2021deepdiagnosis, wardat21DeepLocalize} to extract semantic information from the DNN model. Our feature extraction method reduces the input space dimension while preserving the relevant information for bug localization. During DNN training, there can be many available ``raw'' features such as weights and activation values. In our design, we considered several categories of features. First, we collect values of basic parameters (e.g., weight, gradient, loss, and accuracy) during training. We collect the data at each epoch and form a sequence. See features 1--2 in Table~\ref{feature}. Second, we extract the statistical summary of the selected data using the metrics such as Mean and Median. For example, in Table~\ref{feature}, features 3--11 indicate the use of \textit{mean}, \textit{min}, and \textit{max} for gradient, weight, and bias. Third, we referenced useful DNN bug detection rules developed in the related work, e.g.,~\cite{wardat2021deepdiagnosis} and extract the values used in these rules. See features 12--15 in Table~\ref{feature}. For example, to detect \textit{dead node}, prior works check whether the number of inactive nodes is greater than a threshold. Our technique utilizes the number of inactive nodes per \textit{epoch} as a feature to detect the dead node pattern.

Furthermore, we leverage the features containing relevant information to improve the accuracy of our classifier. As indicated by prior works~\cite{wardat2021deepdiagnosis, ZhangAUTOTRAINER, wardat21DeepLocalize, schoop2021umlaut}, training features indicate the occurrence of symptoms and root causes in buggy programs. 
For example, we apply the mean of the absolute function on gradient values to represent the vanishing gradient feature. This symptom can indicate a problem in the activation function (root cause). We use this data to feed our LSTM model (details in \ref{NetworkBuilding}). Table~\ref{feaures} shows the 15 extracted features from each layer, we build a matrix to represent each feature. In particular, in the feature matrix, rows represent the epoch, and columns represent the descriptive statistics. For example, in the matrix for the weight feature, the first row contains the descriptive statistics (e.g., Mean, Min, Max, Median, STD, etc.) of epoch number 1. The second row has the data for epoch 2 and so on. These matrices serve as sequential input for the LSTML model. Since the models we analyze (e.g., see Table~\ref{tab:benchmark}) can have different numbers of layers, each extracted feature is represented by vectors with different lengths.  
We use \textbf{zero} as padding to ensure consistency among the feature vectors. We replace unknown values (e.g., NaN, INF, or -INF) with unique integer identifiers, as our model receives numerical data in the first layer. 
Finally, we merge all matrices into one. The resulting matrix contains the data of 40 epochs and 683 dimensions for each model. We train the LSTM with the 3900 metrics. 

\subsection{Building Deep Learning Network}

\label{NetworkBuilding}


Recurrent neural networks (RNN) are DL models that support sequential data (i.e., time series and text comments)~\cite{du2017deeplog}. Developers often leverage RNNs to solve problems with temporal data (e.g., natural language processing, image captioning, etc.)~\cite{ciregan2012multi,sutskever2014sequence,hinton2012deep}. RNNs differ from other DL networks (i.e., DNN and CNN) by considering input and output as interdependent. To that end, RNNs save the states of prior inputs in memory, using this information to help determine a given sequence output~\cite{hochreiter1997long}. Nevertheless, RNNs are susceptible to vanishing gradient problems when checking long input sequences~\cite{hochreiter1997long}. Due to the long gap between input and output, this issue prevents an RNN model from learning its data dependencies. The Long Short-Term Memory architecture (LSTM)~\cite{hochreiter1997long} was introduced to solve this dependency problem. Since our approach analyzes long data sequences, we leverage an LSTM model. Our model acts as a classifier to predict bugs in the DL program from its semantic information.

\subsubsection{Building LSTM Model \protect\footnote{Our model available at \protect\url{https://github.com/authorunknown326/Deep4Deep/blob/main/Modeling/Classifier.svg}}}
To build our model, we leverage LSTMs to handle data sequences while learning their semantic information. That is because our problem can be represented as a sequence-to-sequence with varying lengths. We design our model as an encoder-decoder with 14 layers.
The architecture of our model consists of two encoders that receive numerical and text data as input, processing them separately, and a decoder to generate the output. The first 3 layers act as an encoder, with the numerical input first being normalized using the BatchNormalization~\cite{ ioffe2015batch} layer to ensure that all values have similar ranges. Then, the results are fed to two LSTM layers. The second encoder takes the text input after tokenizing it and passes it through an embedding layer converting it into a dense vector representation. This vector representation is processed using two LSTM layers in both encoders. These layers maintain long-term dependencies while encoding the input sequence into a compact representation. To avoid overfitting~\cite{baldi2013understanding}, we concatenate the outputs of the two encoders, passing them to the dropout layer. Next, the data is flattened into a single feature vector. For the decoder stage, we use two sets of LSTM and time-distributed layers, which are used to make predictions. The output layer uses softmax, while the others use the Tanh as activation functions. Finally, we define the model using the model class and compile it with an appropriate loss function and optimizer. We implement our approach using Keras~\cite{Keras}, as it facilitates the building process of neural networks. Our LSTM model is trained using the minibatch stochastic gradient descent (SGD) algorithm~\cite{bottou2010large}, with the Adam optimizer~\cite{kingma2014adam}.

In summary, our encoder-decoder model uses the following layers: BatchNormalization and two LSTM layers work as the first encoder. An embedding layer and two LSTM layers work as the second encoder, then the Concatenation layer, Dropout, Flatten, RepeatVector, LSTM, TimeDistributed, LSTM, and, as output, a TimeDistributed layer, which works as a classifier. For more details on parameter tuning, such as batch sizes and the number of training epochs, please see our availability repository~\cite{Deep4DeepRepo}.

\begin{table}[!htb]
 	\caption{List of Mutation Operators}
 	\label{tab:fix}
    \setlength{\parskip}{0.2cm}
    \setlength{\intextsep}{1cm plus .1cm minus .1cm} 
	\setlength{\belowcaptionskip}{.0cm}
     \scalebox{0.95}{
	\begin{tabular}{|p{7cm}|c|}
 		\hline
		 \multicolumn{1}{|c|}{\textbf{Root Causes}} & \textbf{Label}\\
		\hline\hline
		 Correct Model  & 0\\\hline
		\cellcolor[HTML]{EFEFEF}  Change\_Loss\_Function &\cellcolor[HTML]{EFEFEF} 1\\\hline
		 Change\_Batch\_Size & 2\\\hline
		\cellcolor[HTML]{EFEFEF} Change\_Learning\_Rate &\cellcolor[HTML]{EFEFEF} 3\\\hline
 		 Change\_Activation\_Function & 4\\\hline
 		\cellcolor[HTML]{EFEFEF}Add\_Activation\_Function &\cellcolor[HTML]{EFEFEF} 5\\\hline
		 Remove\_Activation\_Function  & 6\\\hline
		\cellcolor[HTML]{EFEFEF}  Change\_Optimisation\_Function&\cellcolor[HTML]{EFEFEF} 7\\\hline
 	      Change\_Gradient\_Clip & 8\\\hline
 	     \cellcolor[HTML]{EFEFEF} Change\_Weights\_Initialisation &\cellcolor[HTML]{EFEFEF} 9\\\hline
 	      Change\_Dropout\_Rate & 10\\\hline 
 	\end{tabular}
 	\label{tab:operators}
  }
 \end{table}

\begin{table}[!htb]
\centering
\caption{Benchmark Information}
 \scalebox{0.95}{
\begin{tabular}{|llr|r|r|r|}
\hline
\multicolumn{1}{|c|}{\textbf{No}}                         & \multicolumn{1}{c|}{\textbf{Ref}}                                                            & \multicolumn{1}{c|}{\textbf{\# of Layers}} & \multicolumn{1}{c|}{\textbf{Single}} & \multicolumn{1}{c|}{\textbf{Multi}} & \multicolumn{1}{c|}{\textbf{Upvotes}} \\ \hline\hline
\rowcolor[HTML]{EFEFEF} 
\multicolumn{1}{|l|}{\cellcolor[HTML]{EFEFEF}\textbf{1}}  & \multicolumn{1}{l|}{\cellcolor[HTML]{EFEFEF}\textbf{~\cite{Kaggle1}}}  & 5                                      & 30                                   & 336                                 & 153                                   \\ \hline
\multicolumn{1}{|l|}{\textbf{2}}                          & \multicolumn{1}{l|}{\textbf{~\cite{Kaggle2}}}                          & 5                                      & 34                                   & 396                                 & 133                                   \\ \hline
\rowcolor[HTML]{EFEFEF} 
\multicolumn{1}{|l|}{\cellcolor[HTML]{EFEFEF}\textbf{3}}  & \multicolumn{1}{l|}{\cellcolor[HTML]{EFEFEF}\textbf{~\cite{Kaggle3}}}  & 6                                      & 8                                    & 19                                  & 114                                   \\ \hline
\multicolumn{1}{|l|}{\textbf{4}}                          & \multicolumn{1}{l|}{\textbf{~\cite{Kaggle4}}}                          & 10                                     & 31                                   & 319                                 & 79                                    \\ \hline
\rowcolor[HTML]{EFEFEF} 
\multicolumn{1}{|l|}{\cellcolor[HTML]{EFEFEF}\textbf{5}}  & \multicolumn{1}{l|}{\cellcolor[HTML]{EFEFEF}\textbf{~\cite{Kaggle5}}}  & 11                                     & 74                                   & 1611                                & 58                                    \\ \hline
\multicolumn{1}{|l|}{\textbf{6}}                          & \multicolumn{1}{l|}{\textbf{~\cite{Kaggle6}}}                          & 4                                      & 2                                    & 1                                   & 57                                    \\ \hline
\rowcolor[HTML]{EFEFEF} 
\multicolumn{1}{|l|}{\cellcolor[HTML]{EFEFEF}\textbf{7}}  & \multicolumn{1}{l|}{\cellcolor[HTML]{EFEFEF}\textbf{~\cite{Kaggle7}}}  & 5                                      & 27                                   & 268                                 & 54                                    \\ \hline
\multicolumn{1}{|l|}{\textbf{8}}                          & \multicolumn{1}{l|}{\textbf{~\cite{Kaggle8}}}                          & 17                                     & 51                                   & 737                                 & 50                                    \\ \hline
\rowcolor[HTML]{EFEFEF} 
\multicolumn{1}{|l|}{\cellcolor[HTML]{EFEFEF}\textbf{9}}  & \multicolumn{1}{l|}{\cellcolor[HTML]{EFEFEF}\textbf{~\cite{Kaggle9}}}  & 5                                      & 10                                   & 33                                  & 49                                    \\ \hline
\multicolumn{1}{|l|}{\textbf{10}}                         & \multicolumn{1}{l|}{\textbf{~\cite{Kaggle10}}}                         & 3                                      & 17                                   & 113                                 & 45                                    \\ \hline
\rowcolor[HTML]{EFEFEF} 
\multicolumn{1}{|l|}{\cellcolor[HTML]{EFEFEF}\textbf{11}} & \multicolumn{1}{l|}{\cellcolor[HTML]{EFEFEF}\textbf{~\cite{Kaggle11}}} & 3                                      & 4                                    & 5                                   & 39                                    \\ \hline
\multicolumn{1}{|l|}{\textbf{12}}                         & \multicolumn{1}{l|}{\textbf{~\cite{Kaggle12}}}                         & 3                                      & 7                                    & 10                                  & 37                                    \\ \hline
\rowcolor[HTML]{EFEFEF} 
\multicolumn{1}{|l|}{\cellcolor[HTML]{EFEFEF}\textbf{13}} & \multicolumn{1}{l|}{\cellcolor[HTML]{EFEFEF}\textbf{~\cite{Kaggle13}}} & 7                                      & 24                                   & 205                                 & 37                                    \\ \hline
\multicolumn{1}{|l|}{\textbf{14}}                         & \multicolumn{1}{l|}{\textbf{~\cite{Kaggle14}}}                         & 2                                      & 35                                   & 464                                 & 34                                    \\ \hline
\rowcolor[HTML]{EFEFEF} 
\multicolumn{1}{|l|}{\cellcolor[HTML]{EFEFEF}\textbf{15}} & \multicolumn{1}{l|}{\cellcolor[HTML]{EFEFEF}\textbf{~\cite{Kaggle15}}} & 5                                      & 20                                   & 148                                 & 33                                    \\ \hline 
\multicolumn{1}{|l|}{\textbf{16}}                         & \multicolumn{1}{l|}{\textbf{~\cite{Kaggle16}}}                         & 3                                      & 25                                   & 239                                 & 32                                    \\ \hline
\rowcolor[HTML]{EFEFEF} 
\multicolumn{1}{|l|}{\cellcolor[HTML]{EFEFEF}\textbf{17}} & \multicolumn{1}{l|}{\cellcolor[HTML]{EFEFEF}\textbf{~\cite{Kaggle17}}} & 4                                      & 61                                   & 862                                 & 30                                    \\ \hline
\multicolumn{1}{|l|}{\textbf{18}}                         & \multicolumn{1}{l|}{\textbf{~\cite{Kaggle18}}}                         & 5                                      & 41                                   & 588                                 & 29                                    \\ \hline
\rowcolor[HTML]{EFEFEF} 
\multicolumn{1}{|l|}{\cellcolor[HTML]{EFEFEF}\textbf{19}} & \multicolumn{1}{l|}{\cellcolor[HTML]{EFEFEF}\textbf{~\cite{Kaggle19}}} & 3                                      & 10                                   & 31                                  & 28                                    \\ \hline
\multicolumn{1}{|l|}{\textbf{20}}                         & \multicolumn{1}{l|}{\textbf{~\cite{Kaggle20}}}                         & 4                                      & 33                                   & 396                                 & 28                                    \\ \hline
\rowcolor[HTML]{EFEFEF} 
\multicolumn{1}{|l|}{\cellcolor[HTML]{EFEFEF}\textbf{21}} & \multicolumn{1}{l|}{\cellcolor[HTML]{EFEFEF}\textbf{~\cite{Kaggle21}}} & 5                                      & 11                                   & 44                                  & 27                                    \\ \hline
\multicolumn{1}{|l|}{\textbf{22}}                         & \multicolumn{1}{l|}{\textbf{~\cite{Kaggle22}}}                         & 8                                      & 14                                   & 35                                  & 27                                    \\ \hline
\rowcolor[HTML]{EFEFEF} 
\multicolumn{1}{|l|}{\cellcolor[HTML]{EFEFEF}\textbf{23}} & \multicolumn{1}{l|}{\cellcolor[HTML]{EFEFEF}\textbf{~\cite{Kaggle23}}} & 7                                      & 26                                   & 237                                 & 27                                    \\ \hline
\multicolumn{1}{|l|}{\textbf{24}}                         & \multicolumn{1}{l|}{\textbf{~\cite{Kaggle24}}}                         & 5                                      & 29                                   & 321                                 & 27                                    \\ \hline
\rowcolor[HTML]{EFEFEF} 
\multicolumn{1}{|l|}{\cellcolor[HTML]{EFEFEF}\textbf{25}} & \multicolumn{1}{l|}{\cellcolor[HTML]{EFEFEF}\textbf{~\cite{Kaggle25}}} & 5                                      & 35                                   & 318                                 & 27                                    \\ \hline
\multicolumn{1}{|l|}{\textbf{26}}                         & \multicolumn{1}{l|}{\textbf{~\cite{Kaggle26}}}                         & 28                                     & 11                                   & 38                                  & 23                                    \\ \hline
\rowcolor[HTML]{EFEFEF} 
\multicolumn{1}{|l|}{\cellcolor[HTML]{EFEFEF}\textbf{27}} & \multicolumn{1}{l|}{\cellcolor[HTML]{EFEFEF}\textbf{~\cite{Kaggle27}}} & 13                                     & 13                                   & 42                                  & 23                                    \\ \hline
\multicolumn{1}{|l|}{\textbf{28}}                         & \multicolumn{1}{l|}{\textbf{~\cite{Kaggle28}}}                         & 9                                      & 12                                   & 34                                  & 21                                    \\ \hline
\rowcolor[HTML]{EFEFEF} 
\multicolumn{1}{|l|}{\cellcolor[HTML]{EFEFEF}\textbf{29}} & \multicolumn{1}{l|}{\cellcolor[HTML]{EFEFEF}\textbf{~\cite{Kaggle29}}} & 5                                      & 33                                   & 391                                 & 21                                    \\ \hline
\multicolumn{1}{|l|}{\textbf{30}}                         & \multicolumn{1}{l|}{\textbf{~\cite{Kaggle30}}}                         & 7                                      & 17                                   & 103                                 & 15                                    \\ \hline\hline
\rowcolor[HTML]{EFEFEF} 
\multicolumn{3}{|c|}{\cellcolor[HTML]{EFEFEF}\textbf{Total}}                                                                                                                                      & \textbf{745}                         & \textbf{8344}                       & \textbf{}                             \\ \hline
\end{tabular}
\label{tab:benchmark}
}
\end{table}
\section{Evaluation}
\label{sec:EVALUATION}
In the evaluation, we answer the following research questions:
\begin{table*}[!htb]
\caption{Comparing the results from UMLAUT, AUTOTRAINER, DeepLocalize, DeepDiagnosis, DeepFD, and Deep4Deep.}
\scalebox{0.7}{
\begin{tabular}{|cc|cccccc|cccccc|}
\hline
\multicolumn{1}{|c|}{\multirow{2}{*}{\textbf{No}}} & \multirow{2}{*}{\textbf{Post \#}} & \multicolumn{6}{c|}{\textbf{Fault   detection}}                                                                                                                                                                                         & \multicolumn{6}{c|}{\textbf{Fault   diagnosis}}                                                                                                                                                                                         \\ \cline{3-14} 
\multicolumn{1}{|c|}{}                             &                                   & \multicolumn{1}{c|}{\textbf{UMLAUT}} & \multicolumn{1}{c|}{\textbf{AutoTrainer}} & \multicolumn{1}{c|}{\textbf{DeepLocalize}} & \multicolumn{1}{c|}{\textbf{DeepFD}} & \multicolumn{1}{c|}{\textbf{DeepDiag}} & \textbf{Deep4Deep}      & \multicolumn{1}{c|}{\textbf{UMLAUT}} & \multicolumn{1}{c|}{\textbf{AutoTrainer}} & \multicolumn{1}{c|}{\textbf{DeepLocalize}} & \multicolumn{1}{c|}{\textbf{DeepDiag}} & \multicolumn{1}{c|}{\textbf{DeepFD}} & \textbf{Deep4Deep}      \\ \hline
\multicolumn{1}{|c|}{\textbf{1}}                   & \textbf{48385830}                 & \multicolumn{1}{c|}{\checkmark}            & \multicolumn{1}{c|}{\checkmark}                 & \multicolumn{1}{c|}{\checkmark}                  & \multicolumn{1}{c|}{\checkmark}            & \multicolumn{1}{c|}{\checkmark}              & \checkmark                    & \multicolumn{1}{c|}{\checkmark}            & \multicolumn{1}{c|}{\checkmark}                 & \multicolumn{1}{c|}{\checkmark}                  & \multicolumn{1}{c|}{\checkmark}              & \multicolumn{1}{c|}{\checkmark}            & \checkmark                    \\ \hline
\multicolumn{1}{|c|}{\textbf{2}}                   & \textbf{44164749}                 & \multicolumn{1}{c|}{\checkmark}            & \multicolumn{1}{c|}{\xmark}                & \multicolumn{1}{c|}{\checkmark}                  & \multicolumn{1}{c|}{\checkmark}            & \multicolumn{1}{c|}{\xmark}             & \checkmark                    & \multicolumn{1}{c|}{\xmark}           & \multicolumn{1}{c|}{\xmark}                & \multicolumn{1}{c|}{\checkmark}                  & \multicolumn{1}{c|}{\xmark}             & \multicolumn{1}{c|}{\checkmark}            & \checkmark                    \\ \hline
\multicolumn{1}{|c|}{\textbf{3}}                   & \textbf{31556268}                 & \multicolumn{1}{c|}{\checkmark}            & \multicolumn{1}{c|}{\xmark}                & \multicolumn{1}{c|}{\checkmark}                  & \multicolumn{1}{c|}{\checkmark}            & \multicolumn{1}{c|}{\checkmark}              & \checkmark                    & \multicolumn{1}{c|}{\checkmark}            & \multicolumn{1}{c|}{\xmark}                & \multicolumn{1}{c|}{\xmark}                 & \multicolumn{1}{c|}{\checkmark}              & \multicolumn{1}{c|}{\checkmark}            & \checkmark                    \\ \hline
\multicolumn{1}{|c|}{\textbf{6}}                   & \textbf{38648195}                 & \multicolumn{1}{c|}{\checkmark}            & \multicolumn{1}{c|}{\xmark}                & \multicolumn{1}{c|}{\checkmark}                  & \multicolumn{1}{c|}{\xmark}           & \multicolumn{1}{c|}{\checkmark}              & \checkmark                    & \multicolumn{1}{c|}{\xmark}           & \multicolumn{1}{c|}{\xmark}                & \multicolumn{1}{c|}{\checkmark}                  & \multicolumn{1}{c|}{\checkmark}              & \multicolumn{1}{c|}{\xmark}           & \checkmark                    \\ \hline
\multicolumn{1}{|c|}{\textbf{7}}                   & \textbf{33969059}                 & \multicolumn{1}{c|}{\checkmark}            & \multicolumn{1}{c|}{\xmark}                & \multicolumn{1}{c|}{\checkmark}                  & \multicolumn{1}{c|}{\checkmark}            & \multicolumn{1}{c|}{\checkmark}              & \checkmark                    & \multicolumn{1}{c|}{\xmark}           & \multicolumn{1}{c|}{\xmark}                & \multicolumn{1}{c|}{\xmark}                 & \multicolumn{1}{c|}{\checkmark}              & \multicolumn{1}{c|}{\xmark}           & \xmark                   \\ \hline
\multicolumn{1}{|c|}{\textbf{8}}                   & \textbf{55328966}                 & \multicolumn{1}{c|}{\checkmark}            & \multicolumn{1}{c|}{\checkmark}                 & \multicolumn{1}{c|}{\xmark}                 & \multicolumn{1}{c|}{\checkmark}            & \multicolumn{1}{c|}{\checkmark}              & \checkmark                    & \multicolumn{1}{c|}{\checkmark}            & \multicolumn{1}{c|}{\checkmark}                 & \multicolumn{1}{c|}{\xmark}                 & \multicolumn{1}{c|}{\checkmark}              & \multicolumn{1}{c|}{\checkmark}            & \checkmark                    \\ \hline
\multicolumn{1}{|c|}{\textbf{9}}                   & \textbf{34311586}                 & \multicolumn{1}{c|}{\checkmark}            & \multicolumn{1}{c|}{\xmark}                & \multicolumn{1}{c|}{\checkmark}                  & \multicolumn{1}{c|}{\checkmark}            & \multicolumn{1}{c|}{\checkmark}              & \checkmark                    & \multicolumn{1}{c|}{\xmark}           & \multicolumn{1}{c|}{\xmark}                & \multicolumn{1}{c|}{\checkmark}                  & \multicolumn{1}{c|}{\checkmark}              & \multicolumn{1}{c|}{\checkmark}            & \checkmark                    \\ \hline
\multicolumn{1}{|c|}{\textbf{10}}                  & \textbf{31880720}                 & \multicolumn{1}{c|}{\checkmark}            & \multicolumn{1}{c|}{\xmark}                & \multicolumn{1}{c|}{\checkmark}                  & \multicolumn{1}{c|}{\checkmark}            & \multicolumn{1}{c|}{\checkmark}              & \checkmark                    & \multicolumn{1}{c|}{\xmark}           & \multicolumn{1}{c|}{\xmark}                & \multicolumn{1}{c|}{\xmark}                 & \multicolumn{1}{c|}{\checkmark}              & \multicolumn{1}{c|}{\checkmark}            & \xmark                   \\ \hline
\multicolumn{1}{|c|}{\textbf{11}}                  & \textbf{39525358}                 & \multicolumn{1}{c|}{\checkmark}            & \multicolumn{1}{c|}{\xmark}                & \multicolumn{1}{c|}{\checkmark}                  & \multicolumn{1}{c|}{\checkmark}            & \multicolumn{1}{c|}{\xmark}             & \checkmark                    & \multicolumn{1}{c|}{\xmark}           & \multicolumn{1}{c|}{\xmark}                & \multicolumn{1}{c|}{\xmark}                 & \multicolumn{1}{c|}{\xmark}             & \multicolumn{1}{c|}{\checkmark}            & \checkmark                    \\ \hline
\multicolumn{1}{|c|}{\textbf{12}}                  & \textbf{39217567}                 & \multicolumn{1}{c|}{\checkmark}            & \multicolumn{1}{c|}{\xmark}                & \multicolumn{1}{c|}{\xmark}                 & \multicolumn{1}{c|}{\checkmark}            & \multicolumn{1}{c|}{\checkmark}              & \checkmark                    & \multicolumn{1}{c|}{\xmark}           & \multicolumn{1}{c|}{\xmark}                & \multicolumn{1}{c|}{\xmark}                 & \multicolumn{1}{c|}{\xmark}             & \multicolumn{1}{c|}{\xmark}           & \xmark                   \\ \hline
\multicolumn{1}{|c|}{\textbf{13}}                  & \textbf{48934338}                 & \multicolumn{1}{c|}{\checkmark}            & \multicolumn{1}{c|}{\xmark}                & \multicolumn{1}{c|}{\checkmark}                  & \multicolumn{1}{c|}{\checkmark}            & \multicolumn{1}{c|}{\checkmark}              & \checkmark                    & \multicolumn{1}{c|}{\checkmark}            & \multicolumn{1}{c|}{\xmark}                & \multicolumn{1}{c|}{\xmark}                 & \multicolumn{1}{c|}{\xmark}             & \multicolumn{1}{c|}{\checkmark}            & \xmark                   \\ \hline
\multicolumn{1}{|c|}{\textbf{39}}                  & \textbf{37213388}                 & \multicolumn{1}{c|}{\checkmark}            & \multicolumn{1}{c|}{\xmark}                & \multicolumn{1}{c|}{\checkmark}                  & \multicolumn{1}{c|}{\checkmark}            & \multicolumn{1}{c|}{\xmark}             & \checkmark                    & \multicolumn{1}{c|}{\xmark}           & \multicolumn{1}{c|}{\xmark}                & \multicolumn{1}{c|}{\xmark}                 & \multicolumn{1}{c|}{\xmark}             & \multicolumn{1}{c|}{\xmark}           & \xmark                   \\ \hline
\multicolumn{1}{|c|}{\textbf{40}}                  & \textbf{Github\#1}                & \multicolumn{1}{c|}{\checkmark}            & \multicolumn{1}{c|}{\xmark}                & \multicolumn{1}{c|}{\xmark}                 & \multicolumn{1}{c|}{\checkmark}            & \multicolumn{1}{c|}{\checkmark}              & \checkmark                    & \multicolumn{1}{c|}{\xmark}           & \multicolumn{1}{c|}{\xmark}                & \multicolumn{1}{c|}{\xmark}                 & \multicolumn{1}{c|}{\xmark}             & \multicolumn{1}{c|}{\checkmark}            & \checkmark                    \\ \hline
\multicolumn{1}{|c|}{\textbf{41}}                  & \textbf{Github\#2}                & \multicolumn{1}{c|}{\checkmark}            & \multicolumn{1}{c|}{\xmark}                & \multicolumn{1}{c|}{\checkmark}                  & \multicolumn{1}{c|}{\checkmark}            & \multicolumn{1}{c|}{\checkmark}              & \checkmark                    & \multicolumn{1}{c|}{\xmark}           & \multicolumn{1}{c|}{\xmark}                & \multicolumn{1}{c|}{\xmark}                 & \multicolumn{1}{c|}{\checkmark}              & \multicolumn{1}{c|}{\checkmark}            & \xmark                   \\ \hline
\multicolumn{1}{|c|}{\textbf{42}}                  & \textbf{Github\#3}                & \multicolumn{1}{c|}{\checkmark}            & \multicolumn{1}{c|}{\xmark}                & \multicolumn{1}{c|}{\checkmark}                  & \multicolumn{1}{c|}{\xmark}           & \multicolumn{1}{c|}{\checkmark}              & \checkmark                    & \multicolumn{1}{c|}{\xmark}           & \multicolumn{1}{c|}{\xmark}                & \multicolumn{1}{c|}{\xmark}                 & \multicolumn{1}{c|}{\checkmark}              & \multicolumn{1}{c|}{\xmark}           & \xmark                   \\ \hline
\multicolumn{1}{|c|}{\textbf{43}}                  & \textbf{Github\#4}                & \multicolumn{1}{c|}{\checkmark}            & \multicolumn{1}{c|}{\xmark}                & \multicolumn{1}{c|}{\checkmark}                  & \multicolumn{1}{c|}{\checkmark}            & \multicolumn{1}{c|}{\checkmark}              & \checkmark                    & \multicolumn{1}{c|}{\xmark}           & \multicolumn{1}{c|}{\xmark}                & \multicolumn{1}{c|}{\xmark}                 & \multicolumn{1}{c|}{\xmark}             & \multicolumn{1}{c|}{\xmark}           & \xmark                   \\ \hline
\multicolumn{1}{|c|}{\textbf{44}}                  & \textbf{Github\#5}                & \multicolumn{1}{c|}{\checkmark}            & \multicolumn{1}{c|}{\xmark}                & \multicolumn{1}{c|}{\checkmark}                  & \multicolumn{1}{c|}{\xmark}           & \multicolumn{1}{c|}{\checkmark}              & \checkmark                    & \multicolumn{1}{c|}{\xmark}           & \multicolumn{1}{c|}{\xmark}                & \multicolumn{1}{c|}{\checkmark}                  & \multicolumn{1}{c|}{\checkmark}              & \multicolumn{1}{c|}{\xmark}           & \checkmark                    \\ \hline \hline
\multicolumn{2}{|c|}{\textbf{Total}}                                                   & \multicolumn{1}{l|}{44}              & \multicolumn{1}{l|}{9}                    & \multicolumn{1}{l|}{34}                    & \multicolumn{1}{l|}{42}              & \multicolumn{1}{l|}{39}                & \multicolumn{1}{l|}{44} & \multicolumn{1}{l|}{13}              & \multicolumn{1}{l|}{8}                    & \multicolumn{1}{l|}{7}                     & \multicolumn{1}{l|}{26}                & \multicolumn{1}{l|}{24}              & \multicolumn{1}{l|}{24} \\ \hline
\end{tabular}
\label{tab:results}
}
\end{table*}

\begin{table}[!htb]
\caption{Evaluate the performance of our approach }

\scalebox{0.82}{
\begin{tabular}{|c|c|r|r|r|r|r|r|}
\hline
\textbf{\#OP}               & \textbf{Stage}    & \multicolumn{1}{c|}{\textbf{Pre}} & \multicolumn{1}{c|}{\textbf{Rec}} & \multicolumn{1}{c|}{\textbf{Acc}} & \multicolumn{1}{c|}{\textbf{Val\_Pre}} & \multicolumn{1}{c|}{\textbf{Val\_Rec}} & \multicolumn{1}{c|}{\textbf{Val\_Acc}} \\ \hline
\multirow{2}{*}{\textbf{5}} & \textbf{Training} & 0.9388                            & 0.9247                            & 0.9729                            & 0.8741                                 & 0.8681                                 & 0.9486                                 \\ \cline{2-8} 
                            & \textbf{Testing}  & 0.8780                            & 0.8738                            & 0.9505                            & --                                     & --                                     & --                                     \\ \hline
\multirow{2}{*}{\textbf{8}} & \textbf{Training} & 0.9231                            & 0.8177                            & 0.9687                            & 0.7882                                 & 0.7882                                 & 0.9417                                 \\ \cline{2-8} 
                            & \textbf{Testing}  & 0.8683                            & 0.7916                            & 0.9415                            & --                                     & --                                     & --                                     \\ \hline
\end{tabular}
\label{tab:metric}
}
\end{table}

\begin{itemize}
    \item RQ1 (Effectiveness): Does our DL technique localize bugs in Deep Learning programs effectively?
    \item RQ2 (Comparison): How does our approach compare to existing methodologies in terms of time and effectiveness?
    \item RQ3 (Limitation): Which type of bugs does our approach fail to localize?
    \item RQ4 (Ablation): Which features most affect the overall performance of our approach?
    
\end{itemize}

\subsection{Experimental setup}

\subsubsection{Implementation}
We implemented Deep4Deep on top of Keras 2.3.1~\cite{Keras} and TensorFlow 2.1.0~\cite{TensorFlow}. For the feature extraction in Table~\ref{feaures}, we implemented it as a callback method by overriding the method called (on\_epoch\_end(epoch, logs=None)). The overridden method stores the feature in a file as sequence data. Then it passes a sequence of data as training/testing data for the LSTM model. 

We used a computer with a 4 GHz Quad-Core Intel Core i7 Processor and 32 GB 1867 MHz DDR3 GB of RAM running the 64-bit iMac version 4.11.

\begin{figure}[!htb]
	\centering
        \scalebox{1.15}{\includegraphics[width=3.0in,height=3.1in,clip]{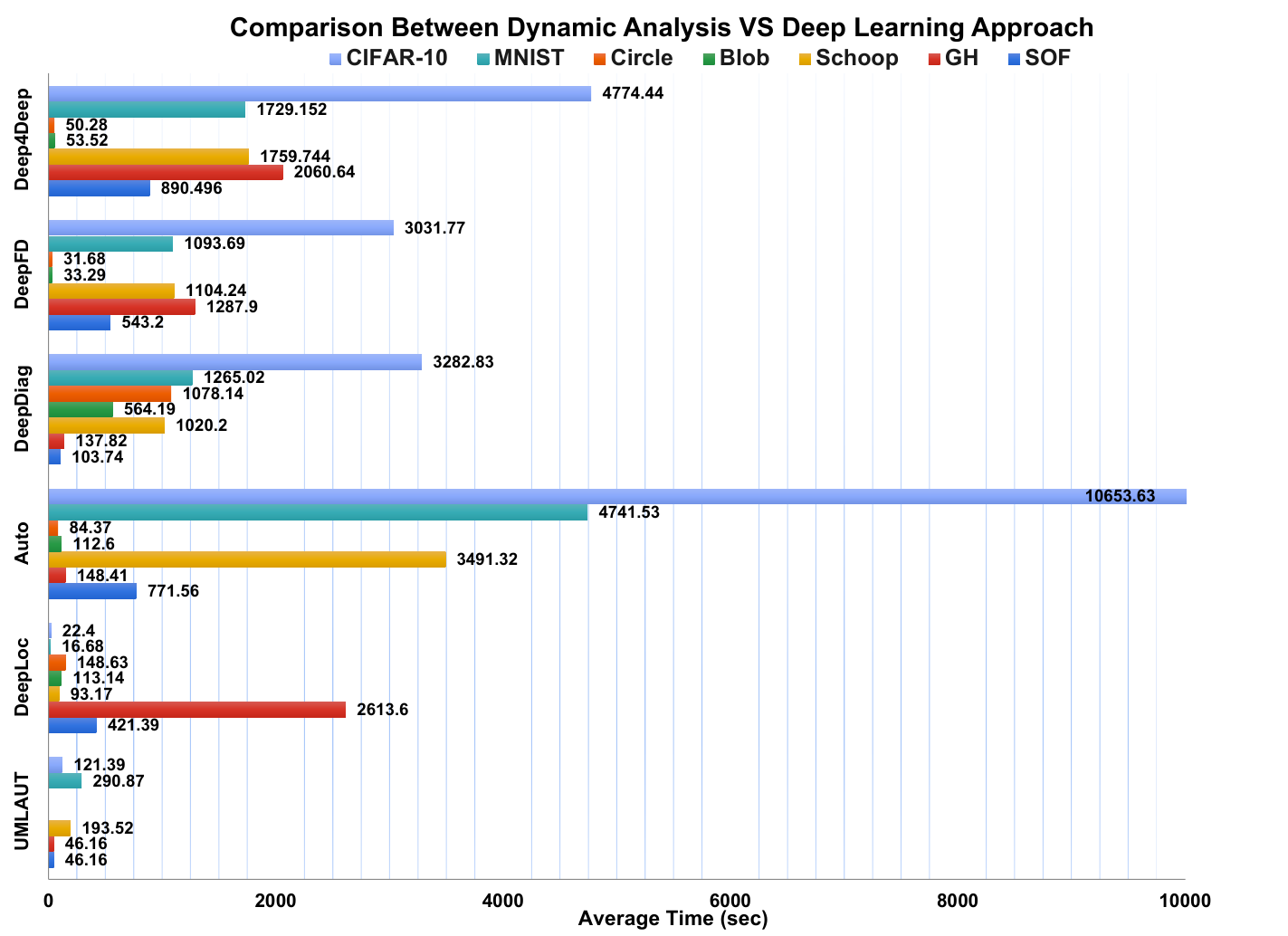}}
	\caption{Comparison between UMLUAT, DeepLocalize (DeepLoc), AUTOTRAINER (Auto), DeepDiagnosis (DeepDiag), DeepFD, and Deep4Deep in terms of seconds.}
	\label{Comparsion}
\end{figure}


\subsection{Preparing the Training Data}
\label{TrainingDataset}
To apply DL for fault localization, it is necessary to collect a large number of correct and buggy DNN programs with their corresponding ground truth labels. This is challenging due to two factors. First, it is difficult to find many buggy models with a training dataset on \sof and \gh. Most models on those platforms have partial code or do not contain bugs related to the DNN source program. Second, to establish the ground truth, it is necessary to verify the fixes to those buggy models. To address this challenge, we follow prior work~\cite{biswas20machine} to generate our training dataset. We collected a set of top-rated DL models from Kaggle~\cite{Kaggle}. We chose Kaggle as it is the most popular platform for hosting and solving data science problems. Users can use Kaggle to find and publish different data sets for solving problems in several domains. We filtered the selected models using the following criteria: First, we selected DNN and CNN models, which our approaches support. Next, we selected the most upvoted Kaggle notebooks. To validate the collected Kaggle models, the first and second authors manually verified each model's accuracy to ensure it is greater than 65\%~\cite{biswas20machine, ZhangAUTOTRAINER}. In the case of multiple models in a Kaggle kernel, we have selected the best-performing one. After applying our filtering criteria, we arrived at 30 correct models. Table~\ref{tab:benchmark} shows the benchmark information, including the reference to the source model, its number of layers, and its number of upvotes. 
\begin{table}[!htb]
\caption{Summary: Bug Detection of SoTA vs. Deep4Deep }

\scalebox{0.73}{
\begin{tabular}{|l|r|r|r|r|r|r|r|}
\hline
                                 & \multicolumn{1}{c|}{\textbf{UM}} & \multicolumn{1}{c|}{\textbf{AT}} & \multicolumn{1}{c|}{\textbf{DL}} & \multicolumn{1}{c|}{\textbf{DD}} & \multicolumn{1}{c|}{\textbf{DFD}} & \multicolumn{1}{c|}{\textbf{D4D}} & \multicolumn{1}{c|}{\textbf{Ground Truth}} \\ \hline
\rowcolor[HTML]{EFEFEF} 
\textbf{Loss\_Function}  & 0 & 0     & 3                    & 9       & 4                                 & 7                                & 15       \\ \hline
\textbf{Batch\_Size} & 0      & 0    & 0     & 0           & 0        & 1                                 & 3                                      \\ \hline
\rowcolor[HTML]{EFEFEF} 
\textbf{Learning\_Rate}  & 3    & 2    & 1       & 6  & 11       & 9                               & 14                                         \\ \hline
\textbf{Activation\_Function}  & 9      & 3       & 3        & 6   & 5     & 4                  & 22                                         \\ \hline
\rowcolor[HTML]{EFEFEF} 
\textbf{Optimisation\_Function}  & 0    & 2     & 0       & 5        & 5               & 4                                & 10                                         \\ \hline
\textbf{Gradient\_Clip}          & 0                                & 0                                & 0                                & 0                                & 0                                 & 0                                 & 0                                          \\ \hline
\rowcolor[HTML]{EFEFEF} 
\textbf{Weights\_Initialisation} & 0                                & 0                                & 0                                & 0                                & 0                                 & 0                                 & 1                                          \\ \hline
\textbf{Dropout\_Rate}           & 0                                & 0                                & 0                                & 0                                & 0                                 & 0                                 & 0                                          \\ \hline
\hline
\rowcolor[HTML]{EFEFEF} 
\textbf{Total}                   & 12                               & 7                                & 7                                & 26                               & 25                                & 25                                & 65                                         \\ \hline
\end{tabular}
\label{tab:number}
}
\end{table}

To automatically generate buggy models, we used DeepCrime~\cite{humbatova2021deepcrime}, a model mutation framework, to create mutant variants of correct models. DeepCrime supports 24 mutation operators out of 35 identified by empirical studies on real faults in DL systems~\cite{humbatova2020taxonomy, islam19, zhang2018empirical}. DeepCrime generates buggy models via mutation, which are similar to real-world ones~\cite{humbatova2021deepcrime}. DeepCrime uses the~\textit{isKilled()} method to verify whether a mutated model is buggy. The output of this mutation tool is a set of reports containing the total mutation score for all operators, p-value, and effect size. We automatically parse this file to classify whether the mutant was killed by DeepCrime, thus indicating a correct bug injection. Due to the stochastic nature of DNN programs, we follow prior works~\cite{ZhangAUTOTRAINER, humbatova2021deepcrime} to reduce the randomness and collect the model's performance accuracy. In particular, we set DeepCrime to retrain each model 5 times.

Table~\ref{tab:operators} shows the 10 DL mutation operators we use to inject structure bugs. We choose these 10 operators due to the time constraints to generate and validate the mutants. In particular, we select the 10 most frequent fault root causes in DL models~\cite{islam19, jahangirova2020empirical, zhang2018empirical}. In the future, we plan to generalize our approach to support all root causes of faults. We evaluated the effectiveness of Deep4Deep on models from GitHub and StackOverflow (SO) containing real faulty models. Our results show a correlation between the detection of mutation-generated and real-world faults. Lastly, Deep4Deep can detect faults with high accuracy in real-world and mutated models (results in Section~\ref{sec:EVALUATION}).

To create a buggy model with multiple faults, we need to identify the \textbf{exact operator}, which causes an issue, and then use a different operator to inject another independent bug into a target model. To that end, we extend DeepCrime~\cite{humbatova2021deepcrime} to reinsert the remaining operators from Table~\ref{tab:operators}, thus automatically generating buggy models with multiple bugs. Algorithm~\ref{alg:Inject-Bug} shows the procedure we use to inject multiple bugs in a DL model. We use this procedure to build the benchmark and ground truth. Then, we feed this dataset to our LSTM model. Algorithm~\ref{alg:Inject-Bug} starts by receiving a correct deep learning (CDL) model as input. Then, the algorithm executes DeepCrime on the CDL. In lines 4-7, the algorithm uses the resulting DeepCrime report (CSV file) to verify which operators introduced bugs to the model, by iterating over each one in the file. Also, the algorithm classifies the bugs based on operator category (e.g., activation or loss function). Next, it combines all operators from different categories (lines 8-12). For example, the algorithm combines the two groups of bugs, one from \textit{loss function} (MSE) and the other from \textit{optimizer} (SGD). Finally, the algorithm returns a set of mutated models, each containing multiple bugs from different groups. 

In summary, to automatically generate the buggy models, we feed the 30 selected models from Kaggle to the Algorithm~\ref{alg:Inject-Bug}. Table~\ref{tab:benchmark} shows a sample of the generated buggy models. Please refer to the availability package for the complete list of models~\cite{Deep4DeepRepo}, which contains 745 models with single bugs, and 8344 with multiple faults.


\begin{algorithm}
	\caption{Inject Multiple Faults}
	\label{alg:Inject-Bug}
	\small
	\DontPrintSemicolon
	\SetKwInOut{Input}{Input}
	\SetKwInOut{Output}{Output}
	
	\Indm\Indmm
	\Input{Correct Deep Learning Model (CDL)}
	\Output{Buggy Deep Learning Model (BDL)}
	\Indp\Indpp
	\BlankLine
	$subset \leftarrow []$\;
	$Mutated\_Model \leftarrow []$\;
	$CSV\_Report = $Run DeepCrime(CBL)$ $\;
	\ForEach{ Category $ \in \mathcal CSV\_Report $}{
		\ForEach{ Operator $ \in \mathcal Category $}{
		        $Bugset \leftarrow All(Category[Operator].Killed == TRUE)$\; 
                $subset[Category].append(Bugset)$\; 
		}
	}
	\ForEach{ G_{1} $ \in \mathcal Category $}{
		\ForEach{ G_{2} $ \in \mathcal Category $}{
    		\If{$G_{1}$ is Not Equal $G_{2}$ }
    	    {
    		        $Bug \leftarrow Combination(subset[G_1], subset[G_2])$\; 
                    $Mutated\_Model.append(Bug)$\;
    		}
    	}
	}
	\Return Mutated\_Model\;
	\BlankLine
	\Indm\Indmm
	\Indp\Indpp
\end{algorithm}
\DecMargin{1.0em}


\subsection{Real-world Benchmark Collection}
To evaluate the effectiveness of Deep4Deep, we ran our approach on two benchmarks. First, we collected 58 models from prior work~\cite{wardat21DeepLocalize, Jialun2022DeepFD}. 
Out of these models, we selected 44 which can be executed by our approach as well as baselines. These 44 models are real-world models from \sof and \gh. 
Second, we randomly selected 3900 mutated models from our benchmark (see~\ref{TrainingDataset}). 
We follow prior work~\cite{wang2016automatically} to split the dataset as follows: 70\% training, 15\% validation, and 15\% testing. These models and our experiment results are publicly available in our repository~\cite{Deep4DeepRepo}.

\subsection{The Presentation of the Results} 
\tabref{tab:results} shows our evaluation results when deploying UMLUAT~\cite{schoop2021umlaut}, DeepLocalize~\cite{wardat21DeepLocalize}, DeepDiagnosis~\cite{wardat2021deepdiagnosis}, AUTOTRAINER~\cite{ZhangAUTOTRAINER}, and our approach Deep4Deep to check the 44 models from \sof and \gh. Also, we added the results of DeepFD~\cite{Jialun2022DeepFD} from their paper. \tabref{tab:results} consists of two parts. Please refer to the reproducibility repository~\cite{Deep4DeepRepo} for the complete table. In {\it Fault detection} we report whether the tool can \textbf{detect the bug}. In {\it Fault diagnosis} we report whether the tool successfully \textbf{identified the bug location}. Table~\ref{tab:number} summarizes the information of Table~\ref{tab:results}. In particular, Table~\ref{tab:number} shows how many bugs each tool can detect for each bug class. Also, it shows the number of bugs in the ground truth. From this table, we can observe that Deep4Deep is supports more bug types than the SoTA.
Table~\ref{tab:metric} shows the performance of Deep4Deep using the mutated models. In particular, the number of operators that Deep4Deep can localize (\textbf{\#OP}). The \textbf{stage} corresponds to when Deep4Deep reports its results (i.e., training or testing). \textbf{Pre}, \textbf{Rec}, and \textbf{Acc} represent precision, recall, and accuracy, respectively. Lastly, \textbf{Val\_Pre}, \textbf{Val\_Rec}, and \textbf{Val\_Acc} correspond to precision, recall, and accuracy for validation, respectively.
Table~\ref{tab:feature} shows the feature engineering of our approach. In particular, \textbf{Feature} represents the statistics we applied to the output of hyperparameters during training. \textbf{Stage} corresponds to when Deep4Deep reports the results. \textbf{Precision}|\textbf{Val\_Precision}, \textbf{Recall}|\textbf{Val\_Recall}, and \textbf{Accuracy}|\textbf{Val\_Accuracy} represent precision, recall, and accuracy for training and validation, respectively. The columns \textbf{\#Fault Dection} and \textbf{\#Fault Diagnosis} show how many bugs Deep4Deep correctly identified and localized, respectively.

\subsection{Effectiveness and Comparison}
\subsubsection{\textbf{RQ1 (Effectiveness)}}
Table~\ref{tab:results} and Table~\ref{tab:metric} show the evaluation results for RQ1 in real-world and mutated models, respectively.

\textbf{Deep4Deep} has correctly identified 44 out of 44 buggy models and correctly reported the fault locations for 24 models listed in Table~\ref{tab:results}. We use three metrics to measure fault localization prediction results: precision, recall, and accuracy. These three metrics are widely adopted to evaluate machine learning techniques~\cite{khoshgoftaar2002tree}. Specifically, we conducted two experiments and trained Deep4Deep for 5 and 8 operators. We collected three metrics during the training and testing stage. Also, we collected validation metrics during training. To reduce the randomness caused by the training process, we repeated the experiment five times and recorded the average performance. As shown in Table~\ref{tab:metric}, for the testing stage, our approach has 86\%, 0.79\%, 0.94\%, for precision, recall, and accuracy, respectively. Our results show that Deep4Deep can correctly identify and localize bugs with high precision and accuracy.

\textbf{DeepFD~\cite{Jialun2022DeepFD}} identified 41 buggy models for the 44 models listed in Table~\ref{tab:results}. Out of these, DeepFD successfully reported diagnosis for 23 models. DeepFD provides three classifiers (K-Nearest Neighbors, Decision Tree, and Random Forest). For mutated models, the average result in the last row shows that the Decision Tree outperforms other classifiers and achieves 79.90\% accuracy in the training stage.  

\textbf{DeepDiagnosis~\cite{wardat2021deepdiagnosis}} has correctly identified 39 out of 44 buggy models. It correctly diagnoses the fault location for 26 models. DeepDiagnosis uses a heuristic approach to detect symptoms. DeepDiagnosis uses a decision tree to map symptoms to their likely causes. DeepDiagnosis has some limitations for specific operators. DeepDiagnosis is limited in terms of scalability, as with the increase in the number of supported bug types, the more difficult it becomes to add new rules. That is because the addition of new rules has to be consistent with already existing ones. Deep4Deep supports more bug types than DeepDiagnosis (e.g., dropout rate). 

\textbf{DeepLocalize~\cite{wardat21DeepLocalize}} identified 34 out of the 44 models and successfully diagnosed 7 models. This is because DeepLocalize only handles the bug related to numerical errors. DeepLocalize cannot provide any suggestions to fix the faults. 

\textbf{AUTOTRAINER~\cite{ZhangAUTOTRAINER}} identified 8 buggy models out of 44 buggy models. Furthermore, AUTOTRAINER was only able to repair 7 models. It considers the model as buggy if its accuracy is less than or equal to the threshold of 60\%. AUTOTRAINER supports only classification models and specific operators (e.g., optimizer and weight initializer). Deep4Deep can handle more varieties of semantically related errors than AUTOTRAINER, as shown in Table~\ref{tab:operators}.

\textbf{UMLAUT~\cite{schoop2021umlaut}} identified 43 faulty models out of the 44  while correctly reporting the fault diagnosis for 13 of them. UMLAUT only supports classification problems and can diagnose issues related to some operators (i.e., dropout rate, missing activation function, and learning rate). Deep4Deep supports additional model types, such as regression and classification. Also, Deep4Deep supports more operators than UMLAUT (i.e., batch size, optimizer, and loss function).

\subsubsection{\textbf{RQ2 (Comparison)}}
\label{sec:DFDVSD4D}
To answer this question, we used multiple datasets from prior works~\cite{schoop2021umlaut, wardat21DeepLocalize, ZhangAUTOTRAINER} to compare our approach with the state-of-the-art models of UMLUAT, DeepLocalize (DeepLoc), AUTOTRAINER (Auto), DeepDiagnosis (DeepDiag), and DeepFD. Deep4Deep requires two steps to predict bugs in each model. First, it runs feature extraction. Then, Deep4Deep runs inference. Deep4Deep's inference process takes less than a second, which we consider negligible. We compare the time for feature extraction of Deep4Deep with one from SOTA. Figure~\ref{Comparsion} shows the average execution time in seconds as follows: UMLUAT (46.16), DeepLoc (421.39), Auto (771.56), DeepDiag (103.74), DeepFD (543.20), and Deep4Deep (890.496) for all the Stack Overflow (SOF) benchmarks. For the GitHub (GH) model, the six approaches require, on average, 46.16, 2613.60, 148.41, 137.82, 1287.90, and 2060.64 seconds, respectively. 

For the Schoop~\textit{et al.}~\cite{schoop2021umlaut} dataset, the six approaches take on average, 193.52, 93.17, 3491.32, 1020.2, 1104.24, and 1759.744 seconds, respectively. For the AUTOTRAINER benchmark, the six approaches require, on average, 206.13, 75.21, 3898.03, 1547.55, 1047.61, and 1651.85 seconds, respectively. Lastly, for all benchmarks, the average time for UMLAUT, DeepLoc, Auto, DeepDiag, DeepFD, and Deep4Deep, 139.62, 489.86, 2857.63, 1064.56, 1017.97, and 1616.90 is seconds. 

Deep4Deep requires more time to extract features than UMLAUT, DeepLoc, and DeepDiag, because it needs sequential data from 40 epochs to run inference. We use 40 epochs in Deep4Deep, as it is the lower bound in the Auto benchmark~\cite{ZhangAUTOTRAINER}. Because of that, our approach is faster than the other SoTA in AT benchmark. Our Deep4Deep is faster to diagnose models requiring a long training time (i.e., with more than 40 epochs) than the other approaches. On the other hand, Deep4Deep is slower than UMLAUT, DeepLoc, and DeepDiag in \sof, \gh, and UMLAUT's benchmarks. To ensure a high-confidence interval for additional bug types, Deep4Deep collects more information than these three approaches during training. These three approaches report results once they detect the first abnormal behavior, while Deep4Deep relies on a confidence interval to reduce false alarms. Therefore, there is a trade-off between confidence (i.e., false alarms) and computational time.

\textbf{Deep4Deep vs. DeepFD~\cite{Jialun2022DeepFD}}
DeepFD does not report precision and recall during the training and testing~\cite{Jialun2022DeepFD}. Table~\ref{tab:metric} shows the training, validation, and testing accuracy for 5 and 8 operators. Deep4Deep outperforms the best classifier from DeepFD~\cite{Jialun2022DeepFD}. Please refer to~\cite{Jialun2022DeepFD} for DeepFD’s evaluation results while analyzing its dataset. ~\tabref{feaures} shows the number of features supported by Deep4Deep. D4D supports 33 features, while DeepFD 20. We show that these features are useful for enabling LSTM models to learn additional patterns and to support new bug types. Deep4Deep supports numerical and structural features, while DeepFD only uses numerical ones. 
DeepFD uses three classifiers~\cite{Jialun2022DeepFD} to localize only 5 fault types. When compared to SoTA, Deep4Deep supports 10 fault types (see ~\tabref{tab:results-operators}). Deep4Deep is trained using all operator ranges supported by DeepCrime, while DeepFD uses a subset of this range~\cite{Jialun2022DeepFD}. Also, our technique is more extensible w.r.t. new bug types than SoTA that is because it leverages an LSTM model, which can learn more complex relationships than ML classifiers~\cite{sutskever2014sequence, vinyals2015show, bahdanau2015neural}. For example, DeepFD uses three ML classifiers that support 20 features. Based on that, a developer can use our checkpoint of the LSTM model and finetuning using a new training dataset with new bug types. \tabref{tab:feature} shows the results of our experiments. In particular, Deep4Deep is not affected by the increasing number of bug types. Please note that while the execution time report is different for DeepFD than the reported in~\cite{Jialun2022DeepFD}, we report the time necessary to train and predict the existence of bugs. DeepFD is faster than Deep4Deep as it requires 20 epochs to run predictions, while Deep4Deep uses 40. We chose 40 epochs, as prior works have confirmed that additional epochs are necessary DNNs to converge and reach a stable state~\cite{glorot2010understanding, ioffe2015batch, zhangfixup}. Our observations indicate that feature extraction should take place at a DNN stable state to improve bug localization.


\subsection{\textbf{RQ3 (Limitation)}}
Our approach failed to diagnose and localize 20 fault models, see Table~\ref{tab:results}. 
Next, we explain a few examples in which Deep4Deep failed to diagnose fault localization. Our technique uses callbacks in Keras, which is inspired by the work of Wardat~\etal~\cite{wardat21DeepLocalize} to extract features. This callback does not yet support models with fit\_generator() instead of the fit() function. Our approach is limited in terms of network structure due to the usage of DeepCrime~\cite{humbatova2021deepcrime} to generate buggy models. DeepCrime only supports hyperparameters related to DNNs. Also, DeepCrime does not support LSTM models. Our approach does not support the detection of bugs related to setting the number of nodes in hidden layers or epochs. Also, our approach does not support faults related to training data (e.g., normalization and balancing). That is because our dataset does not include those types of problems, a typical out-of-the-vocabulary problem for data-driven approaches. We plan to complement our dataset with all these fault types. Also, we intend to use it to improve our approach in future work. 

For example, the following training data-related bugs in Stack Overflow \#52800582, \#44066044 are unsupported by Deep4Deep. In particular, the model training samples are not enough to identify significant features. Also, the developer did not process the training dataset correctly by using one of the following methods -- scaling or normalization. 

Both \#37213388 and \#44998910 programs are related to the epochs faults. Our approach does not support this type of bug because we do not generate models with those issues. Also, we fixed the number of epochs for feature extraction (see~\secref{FeatureExtraction}). We plan to investigate this type of fault in future work. 

Currently, Deep4Deep supports various structures, including convolutional neural networks (CNNs) and fully connected layers. But, Deep2Deep does not support Recurrent Neural Networks (RNNs). Developers can extend our technique to support RNNs and other architectures by collecting or mutating these models and doing feature engineering to extract additional features related to fault for these models and feed them to our  LSTM model. Prior work show has limitation to a specific type of problem. 
For example, UMLUAT and Auto support classification problems. DeepLoc only supports numerical problems, and it does provide any fix suggestions. DeepDiag is a heuristic-based approach for detecting symptoms. It maps a symptom to its root causes using a Decision Tree, which uses a set of rules. To extend this approach to support additional problems, developer needs to add new rules to the Decision Tree. This process is hard as developers have to ensure consistency between the rules in the decision tree. DeepFD uses a machine learning classifier to predict five fault types in DL problems. This approach is not easy to extend to support more types of problems, as ML classifiers do not work well with many features, thus making them more complex. Also, overfitting can easily take place on ML classifier, which can result in low performance result~\cite{novakovic2017evaluation}.


\subsection{\textbf{RQ4 (Ablation)}}
To answer this question, we individually study different aspects of features. Table~\ref{tab:feature} shows which features are used to diagnose faults in the buggy model from the benchmark in Table~\ref{tab:results}. 

Table~\ref{tab:feature} shows the number of identified and diagnosed faults, respectively. We found that by applying the features extracted from the mean, our approach diagnosed 12 faults. Most of these faults are related to batch size operators. Regarding the feature extracted using variance, our approach localized 13 faults. Of these, most of them are related to Learning rate problems. For the features computed using the standard error of the mean (SEM), our technique localized 16 bugs. The majority of those faults are related to activation faults. Regarding features extracted from skewness, our technique localized 18, of which the majority related to the optimizer operator. Our approach diagnosed 19 features reported by the standard deviation. Most of those are related to loss function bugs. Lastly, by combining all the above features, our approach correctly diagnosed 24 faults. Furthermore, by using combined features, our approach achieved the highest precision, recall, and accuracy during the training, validation, and testing stage. 

\begin{table}[!htb]
 \caption{Feature Engineering}
 \centering
 \scalebox{0.80}{
\begin{tabular}{|c|c|r|r|r|r|r|r|r|r|}
\hline
\textbf{\rotatebox[origin=c]{90}{Features} }             & \textbf{\rotatebox[origin=c]{90}{Stage}}    & \multicolumn{1}{c|}{\textbf{\rotatebox[origin=c]{90}{Precision}} }& \multicolumn{1}{c|}{\textbf{\rotatebox[origin=c]{90}{Recall}}} & \multicolumn{1}{c|}{\textbf{\rotatebox[origin=c]{90}{Accuracy}}} & \multicolumn{1}{c|}{\textbf{\rotatebox[origin=c]{90}{Val\_Precision}}} & \multicolumn{1}{c|}{\textbf{\rotatebox[origin=c]{90}{Val\_Recall}}} & \multicolumn{1}{c|}{\textbf{\rotatebox[origin=c]{90}{Val\_Accuracy}}} & \multicolumn{1}{c|}{\textbf{\rotatebox[origin=c]{90}{\# Fault Detection}}} & \multicolumn{1}{c|}{\textbf{\rotatebox[origin=c]{90}{\# Fault Diagnosis}}} \\ \hline
\multirow{2}{*}{\textbf{Mean}} & \textbf{Training} & 0.943                                   & 0.795                                & 0.949                                  & 0.785                                        & 0.759                                     & 0.910                                       & \multirow{2}{*}{44}                              & \multirow{2}{*}{12}                              \\ \cline{2-8}
                               & \textbf{Testing}  & 0.912                                   & 0.821                                & 0.948                                  & --                                           & --                                        & --                                          &                                                  &                                                  \\ \hline
\multirow{2}{*}{\textbf{Var}}  & \textbf{Training} & 0.881                                   & 0.763                                & 0.932                                  & 0.822                                        & 0.746                                     & 0.917                                       & \multirow{2}{*}{44}                              & \multirow{2}{*}{13}                              \\ \cline{2-8}
                               & \textbf{Testing}  & 0.794                                   & 0.704                                & 0.904                                  & --                                           & --                                        & --                                          &                                                  &                                                  \\ \hline
\multirow{2}{*}{\textbf{Sem}}  & \textbf{Training} & 0.914                                   & 0.890                                & 0.961                                  & 0.675                                        & 0.675                                     & 0.870                                       & \multirow{2}{*}{44}                              & \multirow{2}{*}{16}                              \\ \cline{2-8}
                               & \textbf{Testing}  & 0.677                                   & 0.677                                & 0.871                                  & --                                           & --                                        & --                                          &                                                  &                                                  \\ \hline
\multirow{2}{*}{\textbf{Skew}} & \textbf{Training} & 0.952                                   & 0.802                                & 0.952                                  & 0.894                                        & 0.861                                     & 0.952                                       & \multirow{2}{*}{44}                              & \multirow{2}{*}{18}                              \\ \cline{2-8}
                               & \textbf{Testing}  & 0.876                                   & 0.734                                & 0.926                                  & --                                           & --                                        & --                                          &                                                  &                                                  \\ \hline
\multirow{2}{*}{\textbf{STD}}  & \textbf{Training} & 0.944                                   & 0.797                                & 0.950                                  & 0.817                                        & 0.766                                     & 0.919                                       & \multirow{2}{*}{44}                              & \multirow{2}{*}{19}                              \\ \cline{2-8}
                               & \textbf{Testing}  & 0.800                                   & 0.783                                & 0.917                                  & --                                           & --                                        & --                                          &                                                  &                                                  \\ \hline
\multirow{2}{*}{\textbf{All}}  & \textbf{Training} & 0.939                                   & 0.925                                & 0.973                                  & 0.874                                        & 0.874                                     & 0.949                                       & \multirow{2}{*}{44}                              & \multirow{2}{*}{24}                              \\ \cline{2-8}
                               & \textbf{Testing}  & 0.949                                   & 0.874                                & 0.951                                  & --                                           & --                                        & --                                          &                                                  &                                                  \\ \hline
\end{tabular}
\label{tab:feature}}
\end{table}

\subsection{\textbf{Results Discussions}}

\paragraph{Accuracy}DNN fault localization is a non-trivial task~\cite{islam20repairing, islam19}. Our evaluation results show that Deep4Deep can learn bug patterns of a DNN program under analysis. Also, our approach can correctly localize the root causes of at least one bug at a time in those programs. Deep4Deep achieves high precision and recall in our benchmark. Our approach is comparable with prior program analysis techniques (i.e., DeepLocalize, DeepDiagnosis, and AUTOTRAINER). Lastly, our evaluation results show that our feature extraction methodology correctly represents faults in DNN models.

To evaluate our technique, we first generated patches for buggy models in the benchmark. Each patched model resulted in a correct model, which serves as the ground truth. To create the patches, we collected fixes to buggy models from StackOverflow answers with the highest scores and best quality metric. Next, to check for false positives, we ran Deep4Deep on the generated benchmark. The results show that Deep4Deep has a low false positive rate (4 out of 44). For the 4 false positives, Deep4Deep reported messages to change the optimizer and batch size. Upon further investigation, we observed that users prefer to use the SGD optimizer, which takes more iterations than Adam to reach minima. Also, for batch size, since computing the gradient over the entire dataset is expensive, users opted to reduce the batch size to save resources. However, it has been observed in practice that using a larger batch, can lead to a significant improvement in model performance.

\paragraph{Scalability} DeepLocalize, DeepDiagnosis, UMLAUT, and AUTOTRAINER use heuristic-based approaches and rely on hard-coded rules to detect faults. Since these approaches rely on rule-based diagnostic methods, they are limited to the detection of specific faults~\cite{wardat21DeepLocalize, schoop2020scram, wardat2021deepdiagnosis}. Furthermore, it is hard to ensure the scalability and consistency of those approaches with the addition of new rules, as they can interfere with old ones. One of the limitations of ML classifiers is capturing the relationship between the data points when the dataset is very large. Since DeepFD is built using ML classifiers, it would be hard to support new faults. That is not the case for Deep4Deep, as it leverages as an encoder-decoder architecture which can handle high-dimensional time series data.
Lastly, the results of Deep4Deep show that it can detect new bug types without needing rules. For example, Deep4Deep supports batch\_size, dropout\_rate, and weights initializer, which is unsupported by SoTA.

\section[Threats]{Threats to Validity}
\label{sec:THREATSTOVALIDITY}

\paragraph{External Threat} We have collected 30 models from Kaggle, and we used DeepCrime~\cite{humbatova2021deepcrime} to inject bugs and mutate models. To collect quality models, we filtered models with top upvotes with accuracy $>$ 65\% and verified that by executing them manually. Due to randomness in the training model (e.g., initialization of weight values), we may get a different result for the same model. To mitigate the threat, we trained both models, the original model and its mutation, five times to verify that the models have performance issues in all cases. 

\paragraph{Internal Threat} 

When implementing the approach, we used algorithm~\ref{alg:Inject-Bug} to generate the mutation dataset, which we use for training, validation, and testing. To mitigate errors in the approach's implementation that may affect the evaluation and result, we carefully reviewed our implementation to reduce the chances that major errors were compared to the state-of-the-art. Also, we followed prior works~\cite{ZhangAUTOTRAINER, wardat21DeepLocalize, wardat2021deepdiagnosis,schoop2021umlaut} 
to collect symptoms as features and build our model inspired from~\cite{du2017deeplog}. Regarding the dataset, we split it into three parts (i.e., training, validation, and testing). We used the testing split to evaluate our approach. The testing dataset may not represent real-world buggy models. To mitigate this threat, we evaluated Deep4Deep on a set of real-world benchmarks collected from prior works~\cite{Jialun2022DeepFD, wardat21DeepLocalize}.

\section[Related]{Related Work}
\label{sec:relatedwork}
In this section, we detail the related work in fault localization and
bug repair in DL programs. Also, we provide additional details
related to ideas for our technique.

\textbf{Fault localization in Deep Neural Networks:} 
Cao~\textit{et al.}~\cite{Jialun2022DeepFD} propose DeepFD, a learning-based fault localization framework based on three classifiers (i.e., K-Nearest Neighbors, Decision Tree, and Random Forest). DeepFD extracts features during model training to localize faults. Deep4Deep uses static and dynamic analysis to extract features from the models. Our evaluation results show that Deep4Deep outperforms DeepFD in training and testing accuracy. Please refer to Subsection~\ref{sec:DFDVSD4D} for more details. Schoop~\textit{et al.}~\cite{schoop2021umlaut} proposed UMLAUT, a technique that leverages ten heuristics to detect faults in DL models. UMLAUT provides error messages to help developers understand, locate, and fix bugs. DeepLocalize dynamically analyses the evolving trend of weights and metrics during DNN training to localize faults~\cite{wardat21DeepLocalize}. DeepLocalize focuses on identifying single numerical errors. Similarly to DeepLocalize, Deep4Deep collects information from the training stage of the DL model. Our evaluation details the advantages of a data-driven approach in comparison to rule-based techniques in Section~\ref{sec:EVALUATION}.

DEBAR~\cite{zhang2020detecting} is a static analysis technique that detects numerical bugs in DL programs at the model's architecture level. Debar uses static analysis on tensors and numerical values. Yan~\textit{et al.}~\cite{yan2021exposing} proposed GRIST, a dynamic analysis technique built upon DEBAR. Unlike DEBAR, GRIST applies to deep learning frameworks using dynamic computation graphs such as PyTorch. NeuraLint is a model-based validation approach~\cite{nikanjam2021automatic} that detects bugs in fuzzy neural networks. NeuraLint converts a model into an intermediary/meta state. Then, it runs a set of validation rules (via graph transformations). Currently, NeuraLint supports TensorFlow/Keras programs. Deep4Deep uses a DNN model and features collected at training time to localize faults. Furthermore, Deep4Deep can detect more than one bug in a model under analysis.

\textbf{Bug Repair in Deep Neural Networks:} Ma~\textit{et al.}~\cite{ma2018mode} proposed a technique for model debugging, called MODE. MODE identifies a model's faulty neurons through state differential analysis. Then, similarly to regression testing~\cite{duggal2008understanding}, it selects the relevant inputs for neural network training. Essentially, it selects the appropriate subset of training data for bug repair while balancing training time and accuracy. Zhang \textit{et al.} \cite{zhang2019apricot} proposed a technique, called Apricot, for fixing the faulty neurons in neural networks. Apricot generates a set of permutations from a given model. Each model permutation utilizes a subset from the original dataset. Then, Apricot adjusts the weights of the original model according to the deviation between the sub-model, which correctly classifies a test case versus the original model. Apricot is a computationally expensive approach. Deep4Deep addresses structural bugs (e.g., activation functions misusage) while MODE~\cite{ma2018mode} and Apricot~\cite{zhang2019apricot} support training bugs (e.g., insufficient training data). Wardat~\textit{et al.}~\cite{wardat2021deepdiagnosis} proposed a dynamic analysis technique, DeepDiagnosis, for DNN fault localization. DeepDiagnosis detects 8 types of symptoms and suggests actionable fixes to developers. DeepDiagnosis is not limited to numerical errors. Lastly, DeepDiagnosis provides a benchmark with 444 models. Deep4Deep uses a DNN model and features collected at training time to localize faults. While not providing repair suggestions, Deep4Deep directly reports the root causes, which developers can easily repair.


\textbf{Related Ideas:}
Our approach is inspired by previous work that leverages DL models for defect detection and prediction in traditional programs~\cite{wang2016automatically, du2017deeplog}. Also, it is inspired by works that use code features to identify malware~\cite{shibahara2016efficient} and predict defects~\cite{rahman2013and}. Wang~\etal~\cite{wang2016automatically} analyzes semantic features extracted from a target program AST to predict and locate its defects. Du~\etal~\cite{du2017deeplog} proposed DeepLog, an approach that leverages an LSTM model to analyze a target application execution logs. By checking execution logs, it can help developers fix anomalies in their source. Shibahara~\etal~\cite{shibahara2016efficient} uses natural language techniques and a DL model to analyze packages in a network for malware. Deep4Deep dynamically analyses a DNN training features model to predict faults in its source code.
\section[Conclusion]{Conclusions and Future Work}
\label{sec:conclusion}
In this paper, we present Deep4Deep, a data-driven fault localization for deep neural networks. Deep4Deep leverages an LSTM model to automatically identify and localize DNN bugs. Our approach handles additional bug types, which were unsupported by prior dynamic analysis techniques. Our work shows that DL can be used to localize bugs in DNNs. To that end, our approach extracts dynamic and static features from a model under analysis. In particular, Deep4Deep uses dynamic and static analysis to collect model semantic information during training. Both sets of features help localize faults. Developers can easily extend Deep4Deep to support new bug types and DL model architectures (e.g., CNNs and RNNs). Our evaluation results show that Deep4Deep outperforms prior works (i.e., DeepLocalize and UMLAUT).

We have identified the following research directions for future work. First, we plan to expand the generalizability of our prototype to support additional bug types and architectures. Second, we plan to expand our approach to automatically improve the reliability of DL models by repairing multiple faults at a time. Lastly, we plan to extend our technique to other critical domains in SE, such as DNN program testing.

\section[Data Availability]{Data Availability}
\label{sec:data-availability}
We make available on an anonymous GitHub repository~\cite{Deep4DeepRepo} a collection of 30 Kaggle notebooks, \textit{Mutated} programs, \textit{real-world} benchmarks from previous studies, and the results of the evaluation of Deep4Deep. 




\balance
\bibliographystyle{ACM-Reference-Format}
\bibliography{paper}

\end{document}